\documentclass[prd,twocolumn]{revtex4}
\usepackage{bm}
\usepackage{epsfig}
\usepackage{amsmath}
\newcommand{\be}{\begin{equation}}
\newcommand{\ee}{\end{equation}}
\newcommand{\ba}{\begin{eqnarray}}
\newcommand{\ea}{\end{eqnarray}}
\newcommand{\bml}{\begin{mathletters}}
\newcommand{\eml}{\end{mathletters}}
\newcommand{\bes}{\begin{subequations}}
\newcommand{\ees}{\end{subequations}}

\newcommand{\bi}{\begin{itemize}}
\newcommand{\ei}{\end{itemize}}

\begin{document}
\title{Consequences of a Pati-Salam unification of the electroweak-scale active $\nu_R$
model}
\author{P.Q. Hung}
\email[]{pqh@virginia.edu}
\affiliation{Dept. of Physics, University of Virginia, \\
382 McCormick Road, P. O. Box 400714, Charlottesville, Virginia 22904-4714,
USA}

\date{\today}
\begin{abstract}
If right-handed neutrinos are {\em not} singlets under the
electroweak gauge group as it was proposed in a recent model, they can 
acquire electroweak scale masses and are thus accessible experimentally 
in the near future. When this idea is combined with quark-lepton
unification \`{a} la Pati-Salam, one is forced to introduce new
neutral particles which are singlets under the Standard Model (SM).
These ``sterile neutrinos'' which exist in both helicities and which
are different in nature from the popular particles with the same name 
can have their own seesaw with masses in the keV range for the
lighter of the two eigenstates. The
keV sterile neutrinos have been discussed in the literature as warm dark matter 
candidates with wide ranging astrophysical consequences such as structure
formation, supernova asymmetries, pulsar kicks, etc..In addition, the model
contains W-like and Z-like heavy gauge bosons which might be accessible
at the LHC or the ILC. An argument is presented on why, in this model, it is natural
to have four families which can obey existing constraints.
\end{abstract}
\maketitle
\section{Introduction}
The deep question of whether or not the existence of neutrino masses
has anything to do with parity violation in the weak interactions has
been investigated in classic papers of Ref.\cite{goran} on the
Left-Right symmetric model, linking the scale of ``parity violation''
(the mass of the gauge bosons $W_R$, $M_{W_R}$) to the neutrino masses
$m_{\nu}$: $m_{\nu} \rightarrow 0$ as $M_{W_R} \rightarrow \infty$ (the
V-A limit of the weak interactions). This interesting link
stimulates further looks into the meaning of ``parity violation''
and its possible connection to neutrino masses.


The most popular mechanism, the so-called see-saw mechanism \cite{seesaw}, 
for active neutrinos to have a very small mass is to enlarge
the minimal SM by bringing in right-handed neutrinos and to have two 
widely separated masses:
a Dirac mass $m_D$ coming from a lepton-number conserving term in the Lagrangian which should be
much smaller than a Majorana mass $M_R$ coming from a lepton-number violating
term. The nature of these two mass scales is very model-dependent. 
Nothing is known about either $m_D$ or $M_R$
but only about the ratio $m_D^2/M_R$ (the mass of the lighter neutrino) which is
generally thought to be below an eV or so.  

In a generic
seesaw model, the right-handed neutrinos are {\em sterile} i.e. SM singlets. As
a consequence, the Dirac mass $m_D$ is proportional to the electroweak breaking scale,
although its Yukawa coupling is arbitrary. The right-handed neutrino Majorana mass
$M_R$ on the other hand is generally thought to be close to a typical Grand Unified
Theory (GUT) scale, an expectation which is generally based on the embedding of
the SM into a larger gauge group such as e.g. $SO(10)$. 
(Light sterile neutrinos have been considered in models such as \cite{degouvea}.)

If the right-handed neutrinos are {\em not} SM singlets, the situation can
change dramatically. For example, if they are partners in doublets with
right-handed charged leptons- the so-called mirror leptons which
cannot be SM right-handed charged leptons because of neutral current constraints,
$M_R$ is necessary related to $\Lambda_{EW}$ and furthermore they
can contribute to the invisible Z width unless $M_R > M_Z/2$.
This is a model presented in \cite{hung1} where the right-handed
neutrino masses are ``confined'' to a rather ``narrow'' range
$M_Z/2 <M_R <\Lambda_{EW}$. (Another model where there exists
SM non-singlet neutral leptons was discussed in \cite{borut}.)
Electroweak scale SM {\em active} right-handed neutrinos have an appealing 
aspect to them: They can be proved or disproved  at colliders. One
of the characteristic signals is like-sign dilepton events \cite{hung1},
\cite{waiyee,than}.

The model of \cite{hung1} is based on the assumption of the
existence of mirror fermions (both quarks and leptons) in the SM: For every $SU(2)_L$
SM left-handed doublet (e.g.$(\nu,e)_L$) there is a mirror right-handed doublet
(e.g. $(\nu, e^M)_R$) and
for every $SU(2)_L$ right-handed singlet (e.g. $e_R$), there is a mirror {\em left-handed}
singlet (e.g. $e^{M}_L)$). (The idea of mirror fermions have been entertained
in e.g. \cite{goran2}.) 
In \cite{hung2}, contributions of the mirror charged leptons 
to Lepton-Flavor-Violating (LFV) processes such as $\mu \rightarrow e\, \gamma$ and 
$\tau \rightarrow \mu \, \gamma$ were discussed and constraints were
put on couplings which have implications on both future LFV searches
and like-sign dilepton events. (As mentioned in \cite{hung1,hung2}, the
contribution to the S parameter from the mirror fermions (quarks and leptons) can be offset
by the negative contribution to S from Higgs triplets and doublets and from
the Majorana neutrinos \cite{S,hunge6,hungbran}.)

In \cite{hung1}, no attempt was made to unify quarks and leptons. 
One of the nice features of quark-lepton unification is the
group-theoretical explanation of charge quantization. There
are of course several ways to do so and that might involve a single
Grand Unified (GUT) scale or several intermediate scales. One popular intermediate
unification step is the Pati-Salam group $SU(4)_{PS}$ where quarks
and leptons are grouped into the fundamental representations (the lepton
being the fourth ``color'') \cite{PS}. This is particularly useful
if there is some kind of left-right symmetry.
Because our model contains mirror fermions, it turns out that
unification \`{a} la Pati-Salam with the minimal particle content
of \cite{hung1} {\em cannot} be achieved unless one introduces new
neutral leptons which are singlets under the SM: the ``sterile''
neutrinos. To be more precise, the $SU(2)_L$ singlet particle
content of \cite{hung1} is as follows (the notations used are generic): 
$d_R$, $d^{M}_L$, $u_R$, $u^{M}_L$, $e_R$, $e^{M}_L$. It is shown below
that the $SU(4)_{PS}$ quartets can be $(d_R, e_R)$, $(d^{M}_L, e^{M}_L)$.
The $SU(4)_{PS}$ {\em completion} can be made by introducing {\em SM singlets}
$N_{L,R}$ so that one has $(u_R, N_R)$ and $(u^{M}_L, N_L)$. This
is where the sterile neutrinos of both helicities arise in our model.
Unlike a typical model containing sterile neutrinos where there is usually
an intrinsic link- and hence a tight constraint- between the active
and sterile sectors, it will be seen below that this link is at
best very weak in our model and one is not bound by constraints from the
active sector.

The extension of the SM gauge group to accommodate this fermion content
leads back to the old idea of petite unification
as embodied in the group $SU(4)_{PS} \otimes SU(2)_L \otimes
SU(2)_R \otimes SU(2)^{\prime}_L \otimes SU(2)^{\prime}_R$ \cite{put1} 
The new twist in the present work is the identification of the
electroweak $SU(2)$ group as the diagonal subgroup of
$SU(2)_L \otimes SU(2)_R$, differently from \cite{put1}.
Furthermore this extension
leads to the existence of new heavy gauge bosons in the weak sector
which are different from those which have been entertained so far.
In addition, it will be shown that the Pati-Salam unification
scale is much larger than that entertained in \cite{put1,put2}.
The unification is no longer ``petite'' in this new framework.

We will present the construction of this model from the
bottom up point of view. The plan of the paper is as follows.
First, a summary of the model of \cite{hung1} is presented. Second,
we then show that many features of that model can be easily
understood by an extension of the SM gauge group to a partial
unification scheme in the manner of Petite Unification \cite{put1}. 
It will be shown in that section that, in order to complete the
fermion assignment of the model, one has to introduce new
``sterile neutrinos'' which come with both chiralities and
which might have interesting astrophysical consequences.
Third, we discuss the various mass scales including those
that are relevant to those sterile neutrinos, as well as
some possibilities concerning the masses of the mirror fermions.
Fourth, we present
some aspects of phenomenology of the model, including
the computation of $\sin^{2}\theta_{W}(M_Z^2)$ which
is analogous to the one carried out in \cite{put1} for
a similar model. Finally, we end the paper with a brief discussion
on the embedding of the above model into a still larger orthogonal group
in which families from spinors arise in an interesting way. In particular,
we will present an argument supporting the possibility of a fourth
family as well as an argument why the fourth neutrino can have a mass
of the order of the electroweak scale (evading the Z width bound).

\section{Mirror fermions and electroweak scale right-handed neutrinos: A review}
\label{review}

Below we will review the essential elements of the model presented in \cite{hung1}.
We start with a change of notation concerning the weak group $SU(2)_L$. Since
our model contains both left and right-handed fermions transforming
in a similar way under $SU(2)_L$, it would be appropriate to change
$SU(2)_L$ into $SU(2)_V$. 

\bi

\item {\bf Gauge group}: $SU(3)_c \otimes SU(2)_V \otimes U(1)_Y$.

\item {\bf Fermion content}: (Mirror fermions will be accompanied with the superscript $M$
and the subscript $i$ below denotes a family index) 

\bes
\be
\label{sml}
l_L = \left( \begin{array}{c}
\nu_L \\
e_{L}
\end{array} \right)_{i} = (1,2,Y/2=-1/2) \,,
\ee
\be
\label{mirrorl}
l^{M}_R = \left( \begin{array}{c}
\nu^{M}_R \\
e^{M}_{R}
\end{array} \right)_{i} = (1,2,Y/2=-1/2) \,,
\ee
\be
\label{smq}
q_L = \left( \begin{array}{c}
u_L \\
d_{L}
\end{array} \right)_{i} = (3,2,Y/2=1/6) \,,
\ee
\be
\label{mirrorq}
q^{M}_R = \left( \begin{array}{c}
u^{M}_R \\
d^{M}_{R}
\end{array} \right)_{i} = (3,2,Y/2=1/6) \,,
\ee
\be
\label{smlsinglet}
e_{iR}= (1,1,Y/2=-1) \,,
\ee
\be
\label{mirrorlsinglet}
e^{M}_{iL}= (1,1,Y/2=-1) \,,
\ee
\be
\label{smusinglet}
u_{iR}= (3,1,Y/2=2/3) \,,
\ee
\be
\label{mirrorusinglet}
u^{M}_{iL}= (3,1,Y/2=2/3) \,,
\ee
\be
\label{smdsinglet}
d_{iR}= (3,1,Y/2=-1/3) \,,
\ee
\be
\label{mirrordsinglet}
d^{M}_{iL}= (3,1,Y/2=-1/3) \,.
\ee
\ees

\item {\bf Higgs content}:
\be
\label{smhiggs}
\Phi= \left( \begin{array}{c}
\phi^{+} \\
\phi^{0}
\end{array} \right) = (1,2,Y/2=1/2)
\ee
\be
\label{chi}
\tilde{\chi} = \frac{1}{\sqrt{2}}\,\vec{\tau}.\vec{\chi}=
\left( \begin{array}{cc}
\frac{1}{\sqrt{2}}\,\chi^{+} & \chi^{++} \\
\chi^{0} & -\frac{1}{\sqrt{2}}\,\chi^{+}
\end{array} \right) = (1,3,Y/2=1) \,,
\ee
\be
\label{xi}
\xi=\left( \begin{array}{c}
\xi^{+} \\
\xi^{0} \\
\xi^{-}
\end{array} \right) = (1,3,Y/2=0) \,,
\ee
\be
\label{phis}
\phi_S = (1,1,Y/2=0) \,.
\ee

Higgs triplets such as in (\ref{chi}) were considered earlier in various contexts \cite{triplet}.
\item {\bf Majorana mass}

The interaction Lagrangian is listed in \cite{hung1} and, in a more complete form,
in \cite{hung2}. Here, for the review purpose, we will just give the Yukawa interaction
Lagrangian to illustrate the main point of \cite{hung1}: The right-handed neutrino
Majorana mass is of the order of the electroweak scale. 

In what follows, we will assume as in \cite{hung1} a global $U(1)_M$ symmetry
under which the mirror fermions as well as $\tilde{\chi}$ and $\phi_S$ transform
non-trivially while the SM fermions and Higgs doublet are singlets.
This is first to prevent the appearance of a bare Dirac mass term for the neutrinos.
As shown in  \cite{hung1}, it also allows for the coupling giving rise to a
Majorana mass term for the mirror right-handed neutrino while preventing
a Majorana mass term for the left-handed neutrino. Notice that the latter
is known as Type II seesaw where, in a generic scenario, the Higgs triplet
has a small VEV which is not the case here.  
However, it will be seen below
that this is unnecessary when the SM gauge group is extended: Gauge
invariance under the extended gauge group automatically forbids these terms. Furthermore,
the same global $U(1)_M$ symmetry prevents a Majorana mass term for the
left-handed neutrinos. Again, it will be seen below that the Higgs content
of the extended gauge group insures the absence of this term.

The Majorana mass of the right-handed neutrinos comes from the following Lagrangian:
\be
\label{majorana}
{\cal L}_M = l^{M,T}_{R}\, \sigma_2 \,g_{M}\,(\tau_2 \tilde{\chi})\, l^{M}_{R} + H.c.\,.
\ee
Here $l^{M}_{R}$ denotes a column vector with $n$ families to be general, and
$g_{M}$ denotes a $n \times n$ matrix. When
\be
\label{chivev}
\langle \chi^{0}\rangle = v_M \,,
\ee
the Majorana mass matrix for the right-handed neutrinos is
\be
\label{majmatrix}
M_R = g_{M}\,v_M \,.
\ee

\item {\bf Dirac mass}

The Dirac neutrino mass matrix comes from the following Lagrangian
\ba
\label{yuk1}
{\cal L}_S &=& - \bar{l}^{0}_{L}\,g_{Sl}  \, l^{0,M}_{R}\,\phi_S + H.c. 
\nonumber \\
&=& - (\bar{\nu}^{0}_{L}\, g_{Sl}\, \nu^{0,M}_{R} + 
\bar{e}^{0}_{L} \,g_{Sl}\, e^{0,M}_{R})\,\phi_S
+ H.c \,,
\ea
where the superscript $0$ is only relevant for the charged lepton part
where, as shown in \cite{hung2}, the charged lepton mass eigenstates
mainly come from the coupling to $\Phi$. Here $g_{Sl}$ is an $n \times n$ matrix.
With $\langle \phi_S \rangle = v_S$, the Dirac neutrino mass matrix is
\be
\label{dirmatrix}
m_D = g_{Sl} \, v_S \,.
\ee

\item {\bf See-saw}

The see-saw mass matrix is
\be
\label{2x2}
M_2=\left( \begin{array}{cc}
0&m_D \\
m_D&M_R
\end{array} \right) \,.
\ee
In \cite{hung1}, various implications of the model in which $M_R = O(100\,GeV)$
were discussed. In particular, in order for the light neutrino mass to be of $O(<1\,eV)$,
one should have $m_D \sim 10^{5}\,eV$. We will come back to this point below.

\item {\bf Custodial symmetry and $\rho=1$}

When the eigenvalues of $M_R$ are much larger than those of $m_D$, one obtains the
usual seesaw formula for the lighter neutrinos
\be
\label{seesaw}
m_{\nu} = - m_D^{T} M_{R}^{-1} m_{D} \,,
\ee
where as the (almost) right-handed neutrino mass matrix is approximately
$M_R$. As discussed in detail in \cite{hung1}, the VEV of $\chi^{0}$, $v_M$,
can be of order of $\Lambda_{EW} \sim 246\, GeV$ because of
a remnant custodial $SU(2)$ that is guaranteed at tree level that comes from
the breaking of a global $SU(2)_L \otimes SU(2)_R$ down to $SU(2)$ with
$\tilde{\chi}$ and $\xi$ being grouped into its $(3,3)$ representation as
\be
\label{chi2}
\chi = \left( \begin{array}{ccc}
\chi^{0} &\xi^{+}& \chi^{++} \\
\chi^{-} &\xi^{0}&\chi^{+} \\
\chi^{--}&\xi^{-}& \chi^{0*}
\end{array} \right) \,.
\ee
Furthermore, the Higgs doublet can be written as
\be
\label{Phi}
\Phi = \left( \begin{array}{cc}
\phi^{0} & -\phi^{+} \\
\phi^{-} &  \phi^{0,*}
\end{array} \right) \,,
\ee
which transforms as $(2,2)$ under $SU(2)_L \otimes SU(2)_R$.
$SU(2)_L \otimes SU(2)_R \rightarrow SU(2)$ when
\be
\label{chivev2}
\langle \chi \rangle = \left( \begin{array}{ccc}
v_M &0&0 \\
0&v_M&0 \\
0&0&v_M
\end{array} \right) \,,
\ee
and
\be
\label{Phivev}
\langle \Phi \rangle = \left( \begin{array}{cc}
v_2 & 0 \\
0 &  v_2
\end{array} \right) \,.
\ee
A detailed discussion of the minimization of the Higgs potential given in \cite{chanowitz}
indicates that the above VEVs take the form as shown in (\ref{chivev2}) in order to have
a proper vacuum alignment. 

This custodial symmetry guarantees that $\rho=1$ at tree level. In consequence,
$v_M$ can be of order of $\Lambda_{EW} \sim 246\, GeV$. In fact,
$\Lambda_{EW}=v= \sqrt{v_2^2 + 8\,v_M^2} \approx 246\,GeV$ with 
$M_W = g\,v/2$ and $M_Z = M_W/\cos \theta_W$. As discussed in \cite{hung1},
the {\em active} right-handed neutrinos have now electroweak scale masses and can
be produced at colliders. This and other issues are discussed in \cite{hung1,hung2}.
Also, the presence of the mirror charged leptons give rise to LFV processes
$\mu \rightarrow e\, \gamma$ and $\tau \rightarrow \mu\,\gamma$ \cite{hung2}
with implications concerning the search for like-sign dileptons.

\item {\bf Mass scale issues}

Several issues were discussed in \cite{hung1} such as the one concerning the scale of the
Dirac mass matrix. If we denote the largest eigenvalue of $m_D$ by $g_{Sl}^{max}\,v_S$ and
requiring that the light neutrino masses be less than $O(1\,eV)$, it can be seen that
$g_{Sl}^{max}\,v_S \lesssim 10^{5}\,eV$ for $M_R \sim O(\Lambda_{EW})$. For example,
one has $v_S \sim 10^{5}\,eV$ for $g_{Sl}^{max} \sim O(1)$ or $v_S \sim O(\Lambda_{EW})$
for $g_{Sl}^{max} \sim O(10^{-6})$. Some speculation on $v_S$ was discussed in
\cite{hung1}. We will return to this issue below.

Another issue of interest is the usual question of why the mirror quarks and leptons are
heavier than their SM counterparts because they have not yet been observed. This is
a quintessential question that goes to the heart of particle physics: Why do
quarks and leptons have the masses we think they have? Needless to say, there
is at the present time no satisfactory answer to that question although there
exists many models of quark and lepton masses. The same uncertainty applies to
any extension of the SM, in this case the existence of mirror fermions in our model.
However, we shall use, as one example, a particular model of quark masses
and suggest how the disparity in masses in the two sectors may have something
to do with the ratio of two breaking scales, or equivalently the ratio
of the two masses: $M_{LR}$ and $M_{Z}$, where $M_{LR}$ is the scale of
the breaking of $SU(2)_L \otimes SU(2)_R$ to $SU(2)_V$.

\ei

\section{Partial unification of SM and mirror quarks and leptons:
$SU(4)_{PS} \otimes SU(2)_L \otimes
SU(2)_R \otimes SU(2)^{\prime}_L \otimes SU(2)^{\prime}_R$}
\label{putreview}

The possibility of unifying quarks and leptons into a single representation
of some larger gauge group which contains the SM is an old idea 
which still retains its appeal despite the present lack of direct experimental
evidence. Among the many desirable features of the idea of unification
is a rather natural explanation for charge quantization a.k.a the relationship
between quark and lepton charges, or that between quark and lepton hypercharge
quantum numbers. In most cases, this relationship comes from the fact that
the hypercharge generator is proportional to one of the generators of
the unifying group. 

In the SM, one may say that the requirement of anomaly cancellation
automatically determines the hypercharge of, say, the leptons once that of the
quarks is given, or vice versa. However, in our model with mirror fermions,
anomaly cancellation can be realized between left and right-handed leptons
separately, e.g. between $l_L$ and $l^{M}_R$, and similarly between left and
right-handed quarks. As viewed from this angle, one might be tempted to
conclude that charge quantization necessitates some form of unification,
perhaps more so than in the SM.

The path to unification can be very diverse. It can go through
several intermediate energy scales or it can go directly to the unification
scale through a desert in between. In what follows, we shall choose the
former path, namely one in which the SM gauge group is embedded into products
of simple groups which could eventually merge into a simple unifying group.  

\subsection{Description of the model}

In a similar fashion to \cite{put1,put2}, we will assume that
the gauge group $SU(3)_c \otimes SU(2)_V \otimes U(1)_Y$ characterized
by three different gauge couplings $g_3$, $g_2$ and $g^{\prime}$ is
embedded in a group which is characterized by two
different couplings: $G_S(g_S) \otimes G_{W}(g_W)$. Here as in \cite{put1,put2},
we will take $G_S(g_S)$ to be the Pati-Salam gauge group $SU(4)_{PS}$ and 
$G_{W}(g_W)$ to be $SU(2)_L \otimes
SU(2)_R \otimes SU(2)^{\prime}_L \otimes SU(2)^{\prime}_R$.

The pattern of symmetry breaking will be as follows
\begin{equation}
\label{pattern}
G \stackrel{\textstyle M}{\longrightarrow} G_1 
\stackrel{\textstyle \tilde{M}}{\longrightarrow} G_2
\stackrel{\textstyle M_{LR}}{\longrightarrow}  SU(3)_{c} \otimes SU(2)_V \otimes U(1)_Y \,,
\end{equation}
where 
\be
\label{G}
G=SU(4)_{PS} \otimes SU(2)_L \otimes
SU(2)_R \otimes SU(2)^{\prime}_L \otimes SU(2)^{\prime}_R \,,
\ee
\be
\label{G1}
G_1 = SU(3)_{c} \otimes SU(2)_L \otimes
SU(2)_R \otimes SU(2)^{\prime}_L \otimes SU(2)^{\prime}_R \otimes U(1)_{S} \,,
\ee
\be
\label{G2}
G_2 = SU(3)_{c} \otimes SU(2)_L \otimes
SU(2)_R \otimes U(1)_{Y} \,.
\ee
Notice that, in the above, $SU(2)_V$ is a {\em diagonal subgroup} of two of the $SU(2)$'s which
we take to be
\be
\label{LRbreak}
SU(2)_L \otimes SU(2)_R \rightarrow SU(2)_V \,.
\ee

In dealing with fermion assignments, it is useful to write down explicitly the charge
operator as follows
\be
\label{charge}
Q=T_{3V}+\frac{Y}{2} \,,
\ee
where
\be
\label{su2v}
T_{3V} = T_{3L} + T_{3R} \,,
\ee
\be
\label{Y/2p}
\frac{Y}{2} = T^{\prime}_{3L} + T^{\prime}_{3R}+\sqrt{\frac{2}{3}}T_{15} \,,
\ee
and
\be\label{T15}
T_{15}=\frac{1}{2\sqrt{6}}
\left(\begin{array}{cccc}
1   &   &   &  \\
    & 1 &   &   \\
    &   &  1 &   \\
    &   &    & -3 
\end{array}\right).
\end{equation}
The factor $\sqrt{\frac{2}{3}}$ denoted by $C_S$ in \cite{put1,put2} is determined
from the charge structure of fermions. For classification purpose shown below,
we will use (\ref{charge}) and (\ref{Y/2p}). (For the purpose of comparison, one
can consult \cite{put1,put2}.)

What we will show in the next section is the need for introducing electrically neutral,
$SU(2)_V$ singlet left and right-handed fermions, the so-called
``sterile neutrinos'' $N_L$ and $N_R$, in order to be able to complete
the fermion assignment of our model.

\subsection{The need for the ``sterile neutrinos'' $N_L$ and $N_R$
}
What are the fermions listed in the last section that can be grouped into
a quartet of $SU(4)_{PS}$? From (\ref{charge}) and (\ref{Y/2p}), one
can easily obtain the following grouping under $SU(4)_{PS} \otimes SU(2)_L \otimes
SU(2)_R \otimes SU(2)^{\prime}_L \otimes SU(2)^{\prime}_R$:
\be
\label{qlsm}
\Psi_L=(\left( \begin{array}{c}
u_L \\
d_{L}
\end{array} \right)_{i}, \left( \begin{array}{c}
\nu_L \\
e_{L}
\end{array} \right)_{i})= (4,2,1,1,1)\,,
\ee
\be
\label{qlmirror}
\Psi^{M}_R=(\left( \begin{array}{c}
u^{M}_R \\
d^{M}_{R}
\end{array} \right)_{i}, \left( \begin{array}{c}
\nu^{M}_R \\
e^{M}_{R}
\end{array} \right)_{i})= (4,1,2,1,1)\,
\ee
as can be seen from looking at (\ref{sml}), (\ref{smq}), (\ref{mirrorl}) and
(\ref{mirrorq}). According to the symmetry breaking pattern (\ref{LRbreak}),
both SM left-handed doublets and the mirror right-handed counterparts transform
as doublets under $SU(2)_V$ as laid out in the model of \cite{hung1}.

This leaves us with the $SU(2)_V$ singlets: (\ref{smlsinglet}), (\ref{mirrorlsinglet}), 
(\ref{smusinglet}), (\ref{mirrorusinglet}), (\ref{smdsinglet}), and (\ref{mirrordsinglet}).
Here the charge operator takes the form:
\be
\label{Qsinglet}
Q_{singlet} = \frac{Y}{2} = T^{\prime}_{3L} + T^{\prime}_{3R}+\sqrt{\frac{2}{3}}T_{15} \,.
\ee
Noticing that
\be\label{T15p}
\sqrt{\frac{2}{3}}\,T_{15}=
\left(\begin{array}{cccc}
1/6   &   &   &  \\
    & 1/6 &   &   \\
    &   &  1/6 &   \\
    &   &    & -1/2 
\end{array}\right) \,,
\end{equation}
it is easy to see that one can group $e_{iR}$ and $d_{iR}$ into a quartet
of $SU(4)_{PS}$ with $T^{\prime}_{3R} = -1/2$ and $T^{\prime}_{3L}=0$. Similarly,
one can group $e^{M}_{iL}$ and $d^{M}_{iL}$ into a quartet
of $SU(4)_{PS}$ with $T^{\prime}_{3R} = 0$ and $T^{\prime}_{3L}=-1/2$.
As it stands with the particle content of \cite{hung1}, it leaves us with
the ``orphaned'' $u_{iR}$ and $u^{M}_{iL}$ which, in this new scheme,
would have $T^{\prime}_{3R} = +1/2$ and $T^{\prime}_{3L}=+1/2$ respectively.
In order to complete the fermion assignment, we propose to add the
following $SU(2)_V$ electrically {\em neutral} singlets: $N_L$ and $N_R$.
These singlets will be, in our scenario, the ``sterile neutrinos''
which are quite different from those with the same names. Here, these
sterile neutrinos come with both chiralities whereas the common
usage of the adjective ``sterile'' in the literature refers to
mainly right-handed neutrinos. Let us
remind ourselves that the right-handed neutrinos of the model of \cite{hung1}
are parts of doublets of $SU(2)_V$ and therefore are non-sterile.
With these ``sterile neutrinos'', $N_L$ and $N_R$, we can complete the
fermionic assignment of the $SU(2)_V$-singlet sector as
\be
\label{qlrsm}
\Psi_R=(\left( \begin{array}{c}
u_R \\
d_{R}
\end{array} \right)_{i}, \left( \begin{array}{c}
N_R \\
e_{R}
\end{array} \right)_{i})= (4,1,1,1,2)\,,
\ee
\be
\label{qllmirror}
\Psi^{M}_L=(\left( \begin{array}{c}
u^{M}_L \\
d^{M}_{L}
\end{array} \right)_{i}, \left( \begin{array}{c}
N_L \\
e^{M}_{L}
\end{array} \right)_{i})= (4,1,1,2,1)\,.
\ee

The introduction of the ``sterile neutrinos'' $N_L$ and $N_R$,
within the Petite Unification scheme, to complete the fermionic
assignment is a rather surprising and interesting extension
of the model of \cite{hung1}. Let us notice that although this model
contains the same gauge group and the same number of fermion
degrees of freedom as the model of \cite{put1}, the big
difference lies in the interpretation of the meaning
of mirror fermions and in the SM gauge group itself. Here,
the weak $SU(2)$ gauge group comes from the breaking
$SU(2)_L \otimes SU(2)_R \rightarrow SU(2)_V$ whereas
it is the $SU(2)_L$ of $G_W$ in \cite{put1}. The
new interpretation has a number of advantages over the old one
such as the emergence of electroweak scale right-handed neutrinos
with interesting implications \cite{hung1,hung2}, a much larger Petite Unification
scale, and several others that will be discussed below. In addition- and
this is an interesting bonus- there exists ``sterile neutrinos'' which
could have important astrophysical implications to which we will come back in the following
sections.

\section{Masses of $N_L$ and $N_R$, and of mirror fermions}
\label{sterile}

Having introduced the sterile neutral leptons $N_L$ and $N_R$, one would like
to have some hints on the possible mass ranges that one might expect for these
leptons. Closer to home, the mirror fermions are assumed to be
heavy since none has been detected so far. The question is why they
are supposed to be heavier than their SM counterparts. In what follows,
we will try to provide some rationale for this aspect of mirror fermions.

\subsection{Generalized ``see-saw'' involving $N_L$ and $N_R$}

In order to discuss neutrino masses, we need to study the Higgs sector of the model.
In particular, we will concentrate in this section on the Higgs fields that
couple to the fermions. In this section, we will leave out the intergenerational
mass splitting and will focus mainly on the mass differences between quarks and
leptons and, in particular, between the neutral leptons and the rest.

\bi

\item {\bf Dirac mass terms involving $N_L$ and $N_R$}:

In this section, the Dirac masses for the neutral leptons involve mixing
between $\nu_L$ and $N_R$, and $\nu^{M}_R$ and $N_L$. This is different from
the Dirac mixing terms between the $SU(2)_V$-active $\nu_L$ and $\nu^{M}_R$.

For SM fermions, the Dirac mass term would be proportional to the product
\be
\label{SMprod}
\bar{\Psi}_L \times \Psi_R = (1+15,2,1,1,2) \,.
\ee
The Higgs field that can couple to this fermion bilinear can be of two types
under $SU(4)_{PS} \otimes SU(2)_L \otimes
SU(2)_R \otimes SU(2)^{\prime}_L \otimes SU(2)^{\prime}_R$:
$(1,2,1,1,2)$ and/or  $(15,2,1,1,2)$. Note that the VEV of
$(1,2,1,1,2)$ would give equal masses to the SM quarks and leptons
with the same $G_W$ quantum numbers. If the $(15,2,1,1,2)$ is also
present, its VEV would split quark and lepton masses. We have the
following notations
\be
\label{higgssm}
\Phi_{S} = (1,2,1,1,2)\,;\,\Phi_{A} = (15,2,1,1,2) \,. 
\ee
Explicitly, one has
\be
\label{S}
\Phi_{S} = \left( \begin{array}{cc}
\phi^{0}_{S,u} & -\phi^{+}_{S,d} \\
\phi^{-}_{S,u} &  \phi^{0,*}_{S,d}
\end{array} \right) \,.
\ee
and
\be
\label{A}
\Phi_{A} = \phi_{A}^{\beta}\,\frac{\lambda_{\beta}}{2} \,,
\ee
where $\frac{\lambda_{\beta}}{2}$ are the generators of $SU(4)_{PS}$ with
$\beta = 1,..,15$ and $\phi_{A}^{\beta}$ is a $2 \times 2$ matrix of
the form shown in (\ref{S}).

Similarly, for the mirror fermions we would have
\be
\label{mirrorprod}
\bar{\Psi}^{M}_R \times \Psi^{M}_L = (1+15,1,2,2,1) \,.
\ee
The related Higgs fields are
\be
\label{higgsmir}
\Phi^{M}_{S} = (1,1,2,2,1)\,;\,\Phi^{M}_{A} = (15,1,2,2,1) \,. 
\ee
Here, we also have
\be
\label{Smir}
\Phi^{M}_{S} = \left( \begin{array}{cc}
\phi^{0,M}_{S,u} & -\phi^{+,M}_{S,d} \\
\phi^{-,M}_{S,u} &  \phi^{0,*,M}_{S,d}
\end{array} \right) \,.
\ee
and
\be
\label{Amir}
\Phi^{M}_{A} = \phi_{A}^{\beta,M}\,\frac{\lambda_{\beta}}{2} \,,
\ee

In the above the superscripts ``S'' and ``A'' refer to a singlet and an adjoint of
$SU(4)_{PS}$.

Let us define
\be
\label{higgssmtil}
\tilde{\Phi}_{S} = \tau_2\, \Phi^{\ast}_{S}\, \tau_2\,;\,\tilde{\Phi}_{A} = 
\tau_2 \, \Phi^{\ast}_{A}\, \tau_2 \,,
\ee
and
\be
\label{higgsmirtil}
\tilde{\Phi}^{M}_{S} = \tau_2\, \Phi^{\ast,M}_{S}\, \tau_2\,;\,\tilde{\Phi}^{M}_{A} = 
\tau_2 \, \Phi^{\ast,M}_{A}\, \tau_2 \,,
\ee

One can now write down the Yukawa interactions which give rise to the Dirac masses as follows.
\ba
\label{Dirac}
{\cal L}_{Dirac}&=& y_{1}\,\bar{\Psi}_L \, \Phi_{S} \, \Psi_R + y_{2}\,
\bar{\Psi}_L \, \tilde{\Phi}_{S} \, \Psi_R + \nonumber \\
&&y_{3}\,\bar{\Psi}_L \, \Phi_{A} \, \Psi_R
+ y_{4}\, \bar{\Psi}_L \, \tilde{\Phi}_{A} \, \Psi_R + \nonumber \\
&&y^{M}_{1}\,\bar{\Psi}^{M}_R \, \Phi^{M}_{S} \, \Psi^{M}_L + y^{M}_{2}\,
\bar{\Psi}^{M}_R \, \tilde{\Phi}^{M}_{S} \, \Psi^{M}_L + \nonumber \\
&&y^{M}_{3}\,\bar{\Psi}^{M}_R \, \Phi^{M}_{A} \, \Psi^{M}_L
+ y^{M}_{4}\, \bar{\Psi}^{M}_R \, \tilde{\Phi}^{M}_{A} \, \Psi^{M}_L  \nonumber \\
&&+H.c.
\ea

We assume the following VEVs:
\be
\label{vevsm}
\langle \phi^{0}_{S,u} \rangle= v_{u}\,;\,
\langle \phi^{0}_{S,d} \rangle= v_{d} \,,
\ee
\be
\label{vevmir}
\langle \phi^{0,M}_{S,u} \rangle= v^{M}_{u}\,;\,
\langle \phi^{0,M}_{S,d} \rangle= v^{M}_{d} \,.
\ee
Since $15 = 8 + 3 + \bar{3} + 1$ under the subgroup $SU(3)_c$, one
can only have
\be
\label{15}
\langle \Phi_{A}  \rangle = \langle \phi_{A}^{15} \rangle \,\frac{\lambda_{15}}{2}\,;\,
\langle \Phi^{M}_{A}  \rangle = \langle \phi^{M,15}_{A} \rangle \,\frac{\lambda_{15}}{2} \,.
\ee
With
\bes
\label{vev15}
\be
\frac{\langle \phi_{A,u}^{15} \rangle}{2\,\sqrt{6}}= v_{15,u} \,;\,
\frac{\langle \phi_{A,d}^{15} \rangle}{2\,\sqrt{6}}= v_{15,d} \,,
\ee
\be
\frac{\langle \phi^{M,15}_{A,u} \rangle}{2\,\sqrt{6}}= v^{M}_{15,u} \,; \,
\frac{\langle \phi^{M,15}_{A,d} \rangle}{2\,\sqrt{6}}= v^{M}_{15,d} \,,
\ee
\ees
we obtain the following mass scales for the SM fermions
\bes
\label{upmass}
\be
\label{up}
m_U = y_{1} v_{u} + y_{2} v_{d} + y_{3} v_{15,u} + y_{4} v_{15,d} \,,
\ee
\be
\label{nu}
m_{\nu_L N_R} = y_{1} v_{u} + y_{2} v_{d} - 3(y_{3} v_{15,u} + y_{4} v_{15,d}) \,,
\ee
\ees
\bes
\label{downmass}
\be
\label{down}
m_D = y_{1} v_{d} + y_{2} v_{u} + y_{3} v_{15,d} + y_{4} v_{15,u} \,,
\ee
\be
\label{e}
m_E = y_{1} v_{d} + y_{2} v_{u} - 3( y_{3} v_{15,d} + y_{4} v_{15,u}) \,,
\ee
\ees
and, for the mirror fermions,
\bes
\label{upmirmass}
\be
\label{upm}
m_{U^M} = y^{M}_{1} v^{M}_{u} + y^{M}_{2} v^{M}_{d} + y^{M}_{3} v^{M}_{15,u} + y^{M}_{4} v^{M}_{15,d} \,,
\ee
\be
\label{numir}
m_{\nu^{M}_R  N_L} =  y^{M}_{1} v^{M}_{u} + y^{M}_{2} v^{M}_{d} -3 ( y^{M}_{3} v^{M}_{15,u} + y^{M}_{4} v^{M}_{15,d})\,,
\ee
\ees
\bes
\label{downmirmass}
\be
\label{downmir}
m_{D^M} =  y^{M}_{1} v^{M}_{d} + y^{M}_{2} v^{M}_{u} + y^{M}_{3} v^{M}_{15,d} + y^{M}_{4} v^{M}_{15,u}\,,
\ee
\be
\label{em}
m_{E^M} = y^{M}_{1} v^{M}_{d} + y^{M}_{2} v^{M}_{u} - 3 ( y^{M}_{3} v^{M}_{15,d} + y^{M}_{4} v^{M}_{15,u}) \,.
\ee
\ees

As we shall see below, the generalized ``see-saw'' involving $\nu_L$, $\nu^{M}_R$, $N_L$, and
$N_R$ will now also depend on $m_{\nu_L N_R}$ and $m_{\nu^{M}_R  N_L}$. Furthermore, it will be seen that,
with the existence of electroweak-scale right-handed neutrinos in our model, any additional ``Dirac'' term
will be constrained to be small (how small this is will be the subject of the next section). There might
be several ways to achieve this and we will show two of such possibilities.

\item {\bf Majorana mass terms involving $\nu^{M}_{R}$}: 

The electroweak-scale right-handed neutrino model of \cite{hung1} invokes two Higgs triplets
(\ref{chi}) and (\ref{xi}). As mentioned above, the Majorana mass term involves (\ref{chi})
while $\rho=1$ at tree-level requires the addition of the triplet (\ref{xi}). In the context of
$SU(4)_{PS} \otimes SU(2)_L \otimes
SU(2)_R \otimes SU(2)^{\prime}_L \otimes SU(2)^{\prime}_R$,
the fermion bilinear of interest would be
\be
\label{putmaj}
\Psi^{M,T}_R \, \sigma_2 \, \Psi^{M}_R= (4 \times 4= 6+10, 1, 1+3, 1, 1) \,.
\ee
Under the subgroup $SU(3)_c$, one has the decompositions: $6= 3 + \bar{3}$ and
$10 = 1 + 3 + 6$. Since one is looking for a Higgs field that has a non-zero VEV,
this Higgs field should contain  a {\em singlet} under $SU(3)_c$ as well as being
a SM triplet. In consequence,
we shall take
\be
\label{10}
\Phi_{10} = (\bar{10}= 1 + \bar{3} + \bar{6}, 1, 3, 1, 1) \,,
\ee
where the $SU(3)_c$ singlet part $(1,1,3,1,1)$ couples only to the leptons..  
The Lagrangian is
\be
\label{majlagrangian}
\tilde{{\cal L}}_{M}= \Psi^{M,T}_R \, \sigma_2 \,g_{M}\,(\tau_2\,\Phi_{10}) \Psi^{M}_R \,.
\ee

The VEV of $\Phi_{10}$ is given by
\be
\label{vevphi10}
\langle \Phi_{10} \rangle= \langle (1,1,3,1,1) \rangle = v_{M} \,.
\ee
Notice from (\ref{chi}), (\ref{chivev}), (\ref{10}) and (\ref{majlagrangian})
that (\ref{vevphi10}) which involves an $SU(3)_c$ singlet only gives
a Majorana mass to the right-handed neutrinos as in (\ref{majmatrix}),
namely $M_R = g_M\,v_M$.

One important remark is in order at this point. In \cite{hung1}, in order
to avoid a Majorana mass term of the type $\nu^{T}_L \sigma_2 \nu_L$ which
could, in principle, come from the coupling to the triplet $\tilde{\chi}$,
a global symmetry $U(1)_M$ was imposed. In the present framework, one notices
that $\Psi^{T}_L \sigma_2 \Psi_L = (6+10, 1+3, 1,1,1)$ does not couple to
$\Phi_{10}$. In consequence, the Majorana mass term does not exist as long as
{\em only} $\Phi_{10}$ is present and there is no need to invoke the $U(1)_M$
symmetry.

The phenomenology of $\Phi_{10}$ is quite interesting involving in particular
its color-non-singlet components. This will be discussed in the phenomenology
section below.

\item {\bf Dirac mass terms involving $\nu_L$ and $\nu^{M}_R$}:
\label{activedirac}

What would be the equivalent of Eq. (\ref{yuk1}) for the neutrino Dirac masses
and how does the singlet scalar field $\phi_S$?

Let us look at the following bilinear:
\be
\label{diracput}
\bar{\Psi}_L \, \Psi^{M}_R = (1+15, 2, 2, 1, 1) \,.
\ee
From (\ref{diracput}), it is clear that a bare Dirac mass term
of the type $m_D\,\bar{\Psi}_L \, \Psi^{M}_R$ is {\em not allowed}
by gauge invariance. In \cite{hung1}, a global $U(1)_M$ symmetry
was imposed by hand to prevent such a term and we have just seen that
it is not necessary to do so here.

Let us choose the following Higgs field with four {\em real} components
\be
\label{putsinglet}
\tilde{\Phi}_S=(1,2,2,1,1) \,.
\ee
How is $\tilde{\Phi}_S$ a singlet like $\phi_S$ under $SU(2)_V$? Obviously, it is
not but one of its components {\em is}. To see this, let us recall
that $SU(2) \otimes SU(2) \sim SO(4)$ and a $(2,2)$ of $SU(2) \otimes SU(2)$
is just a quartet representation of $SO(4)$. We can write
\be
\label{O4}
\tilde{\Phi}_S = (\phi_S, \vec{\pi}_S) \in SO(4) \,,
\ee
where $\vec{\pi}_S$ has three components. Under the diagonal
subgroup $SU(2)_V$ of $SO(4)$, $\vec{\pi}_S$ is a ``vector'' while
$\phi_S$ is a singlet. In summary, {\em under} $SU(2)_V$:
\be
\label{sigma}
\vec{\pi}_S \sim {\bf 3}\,;\,\phi_S \sim {\bf 1} \,.
\ee
Explicitly, we have
\be
\label{Stilde}
\tilde{\Phi}_{S} = \left( \begin{array}{cc}
\phi_S +\imath \,\pi^{3}_{S} & -\frac{1}{\sqrt{2}}(\pi^{1}_{S} + \imath\, \pi^{2}_{S}) \\
 \frac{1}{\sqrt{2}}(\pi^{1}_{S} - \imath \,\pi^{2}_{S})&  \phi_S -\imath \, \pi^{3}_{S}
\end{array} \right) \,.
\ee

A few comments are in order at this point. A quick glance at (\ref{Stilde}) reveals
a form that looks very much like how one would also present a complex Higgs doublet
for the SM $SU(2)_L$. The big difference lies however in the fact that $\phi_S$
in (\ref{Stilde}) is now actually a {\em singlet} of $SU(2)_V$ unlike
the case with the SM. Therefore, its VEV {\em does not} break $SU(2)_V$ as in \cite{hung1}.

The Lagrangian here is
\be
\label{dirlagrangian}
\tilde{{\cal L}}_{D}=\bar{\Psi}_L \,g_{Sl}\,\tilde{\Phi}_{S}\,  \Psi^{M}_R + H.c.
\ee
With $\langle \phi_S \rangle = v_S$, we have
\be
\label{Stildevev}
\langle \tilde{\Phi}_{S} \rangle = \left( \begin{array}{cc}
v_S & 0 \\
 0&  v_S
\end{array} \right) \,.
\ee
(\ref{Stildevev}) gives rise to the Dirac neutrino mass matrix among
$\nu_L$ and $\nu^{M}_R$ as (\ref{dirmatrix}), namely
$m_D = g_{Sl} \, v_S$. Furthermore, (\ref{dirlagrangian}) also gives
rise to mass mixing between SM charged fermions and their mirror fermion
counterparts. This has been thoroughly discussed in \cite{hung1, hung2}
and it was found that those mixings are negligible, for the 
changes to the eigenvalues of the charged fermions are of the form 
$m_D^2/(m_{f^M}-m_{f^{SM}}) \ll (m_{f^M}, \, m_{f^{SM}})$
since, as we briefly review in (\ref{review}), $m_D \sim 10^{5}\,eV$.

At this point, an important remark is in order here. Let us notice
that $\langle \tilde{\Phi}_{S} \rangle=v_S \neq 0$ spontaneously break
$SU(2)_L \otimes SU(2)_R$ down to $SU(2)_V$. As discussed below,
the scale $M_{LR}$ associated with $SU(2)_L \otimes SU(2)_R \rightarrow SU(2)_V$
is of $O(\leq 1\,TeV)$. It is natural to ask whether or not one could identify
$v_S$ with $M_{LR}$. If this is the case, the Yukawa coupling
$g_{Sl}$ would have to be $g_{Sl} \sim 10^{-7}$ in order for
$m_D \sim 10^{5}\,eV$ in an electroweak scale see-saw scenario. There
might not be anything unnatural about the smallness of this Yukawa
coupling since the SM contains couplings of that order such as the
electron Yukawa coupling. $g_{Sl}$ needs not necessarily be of order
unity. This might be the simplest scenario. Furthermore, as we
shall see in the last section of the paper on families from spinors,
it is quite natural in this framework to have a fourth family. There
we argue that the fourth neutrino can be heavy ($> M_Z/2$) while
the other three are light because $g_{Sl}$ is generated at the one-loop level
and can be $\sim 10^{-7}$.
Another more complicated scenario
with $g_{Sl} \sim O(1)$ and $v_S \sim O(10^5\,eV)$
is to add another similar Higgs field with a large VEV and forbids a
coupling of the type (\ref{dirlagrangian}) by some global or discrete symmetry.

\item {\bf Majorana mass terms involving $N_{R}$}

The appropriate fermion bilinear is
\be
\label{putmaj2}
\Psi^{T}_R \, \sigma_2 \, \Psi_R= (4 \times 4= 6+10, 1, 1, 1, 1+3) \,.
\ee
In a similar fashion to (\ref{10}), we introduce a Higgs field
\be
\label{10N}
\Phi_{10N} = (\bar{10}= 1 + \bar{3} + \bar{6}, 1, 1, 1, 3) \,,
\ee
with a Lagrangian
\be
\label{majlagrangianN}
\tilde{{\cal L}}_{M,N}= \Psi^{T}_R \, \sigma_2 \,g_{M,N}\,(\tau_2\,\Phi_{10N}) \Psi_R + H.c.\,.
\ee
Once more we notice that, as long as only $\Phi_{10N}$ is introduced, there
is no coupling of the bilinear $\Psi^{M,T}_L \sigma_2 \Psi^{M}_L$ to $\Phi_{10N}$
which, if present, would give rise to a Majorana mass term $N^{T}_L \sigma_2 N_L$.

The VEV of $\Phi_{10N}$ will be
\be
\label{vevphi10N}
\langle \Phi_{10N} \rangle= \langle (1,1,1,1,3) \rangle = v_{M,N} \,,
\ee
giving rise to the Majorana mass for the right-handed sterile neutrino $N_R$:
\be
\label{majmassNR}
M^{N}_R= g_{M,N}\,v_{M,N} \,.
\ee

\item {\bf Dirac mass terms between $N_L$ and $N_R$}
\label{steriledirac}

The Dirac mass term $\bar{N}_R\,N_L$ is contained in
\be
\label{dirnlnr}
\bar{\Psi}_R \times \Psi^{M}_L = (1+15,1,1,2,2) \,.
\ee
Let us notice that because of (\ref{dirnlnr}), there is {\em no bare
mass} term $\bar{N}_R\, N_L$ since it is {\em forbidden} by gauge invariance.
(\ref{dirnlnr}), in addition to $\bar{N}_R\,N_L$, contains mixings between
the right-handed SM charged fermions with the left-handed mirror fermions,
all of which are $SU(2)_V$ singlets. One can choose for the Higgs field
\be
\label{phiNS}
\Phi^{N}_S = (1,1,1,2,2) \,.
\ee
This Higgs field is of course $SU(2)_V$ singlet. This is similar to the case
considered in \cite{hung1,hung2}. Its VEV is
\be
\label{phiNvev}
\langle \Phi^{N}_S \rangle = \left( \begin{array}{cc}
v^{N}_S & 0 \\
 0&  v^{N}_S
\end{array} \right) \,.
\ee
The relevant Lagrangian is
\be
\label{Nlagrangian}
{\cal L}^{N}_{D}= \bar{\Psi}_R \,g^{N}_{S}\,\Phi^{N}_S  \, \Psi^{M}_L + H.c. \,,
\ee
giving the following Dirac mass
\be
\label{dirN}
m^{N}_D = g^{N}_{S} v^{N}_S \,.
\ee

Once again, a remark concerning the size of $v^{N}_S$ is in order here. If
we assume that $v^{N}_S \sim v_S \sim O(\leq 1\,TeV)$, it follows, from
the discussion presented below, that $g^{N}_{S} \sim 10^{-5}-10^{-7}$. The remarks
made above concerning small values of the Yukawa couplings apply equally to this case.
Let us also notice that the symmetry breaking pattern, at $\tilde{M}$,
$SU(2)^{\prime}_L \otimes SU(2)^{\prime}_R \otimes U(1)_S \rightarrow U(1)_Y$ requires a Higgs
field with non-vanishing $U(1)_S$ quantum number which will not
couple to $\bar{N}_L\,N_R$ for the latter has a vanishing $U(1)_S$ quantum number.
One will not have to worry about the scale $\tilde{M}$ (to be discussed below)

\item {\bf Generalized ``see-saw''}

One can now put the pieces obtained above to write a ``see-saw'' matrix
which is now a $4 \times 4$ matrix instead of $2 \times 2$ one as follows.
\be
\label{4x4}
M_4=\left( \begin{array}{cccc}
0&m_D&0&m_{\nu_L N_R} \\
m_D&M_R&m_{\nu^{M}_R  N_L}&0 \\
0&m_{\nu^{M}_R  N_L}&0&m^{N}_D \\
m_{\nu_L N_R}&0&m^{N}_D&M^{N}_R
\end{array} \right) \,.
\ee

Let us notice that (\ref{4x4}) would decompose into two $2 \times 2$ blocs
had $m_{\nu_L N_R} = m_{\nu^{M}_R  N_L} = 0$, and its diagonalization
becomes straightforward. For the upper $2 \times 2$ bloc, a previous discussion
made in \cite{hung1} implied that, for $M_R \sim O(100\,GeV)$, $m_D \sim 10^{5}\,eV$.
The other elements of $M_4$, namely $m_{\nu_L N_R}, m_{\nu^{M}_R  N_L}, m^{N}_D, M^{N}_R$,
are ``unconstrained'' at this stage. However, they are actually ``constrained'' in the sense that
they might influence the active-sterile mixing angles as we shall see below. 

\item {\bf Numerical examples}

To gain some insight,
let us vary $m_{\nu_L N_R}, m_{\nu^{M}_R  N_L}, m^{N}_D, M^{N}_R$
and observe the pattern of mass eigenvalues. An exhaustive numerical study is beyond the scope
of this paper. However it will be useful to show  a few numerical examples for illustration.

In the discussion that follows we will concentrate on the overall mass scales of
various neutral lepton sectors and will ignore flavor differences for the time being.
Several special cases will now be listed.

\bi

\item  ${\bf m_{\nu_L N_R}= m_{\nu^{M}_R  N_L}=0}$:

There is, of course, no reason why this should be the case but, for the
sake of clarity, it will be shown as a first step in our discussion. The
mass eigenvalues are now straightforwardly given as
\be
\label{ss1}
\frac{M_R \pm \sqrt{M_{R}^2 + 4\, m_{D}^2}}{2} \approx \left\{ 
\begin{array}{ll}
M_R \\
-\frac{m_D^2}{M_R}
\end{array}
\right. \,,
\ee
for the non-sterile sector, where the approximation comes from the case where $m_D \ll M_R$, and
\be
\label{ss2}
\frac{M^{N}_R \pm \sqrt{(M^{N}_{R})^2 + 4\, (m^{N}_{D})^2}}{2} \,,
\ee
for the sterile sector, where Eq. (\ref{ss2}) is left in its full form.

For (\ref{ss1}), we have seen in \cite{hung1} that $m_D \sim 10^{5}\,eV$ for
$M_R \sim O(100\,GeV)$ in order for the light active neutrinos to have masses of
$O(<1\,eV)$. The same thing cannot be said about the sterile masses, although
there appears to be interesting mass ranges in the keV region which might be
of astrophysical interest. We shall come back to this aspect below.

\item  ${\bf m_{\nu_L N_R}= m_{\nu^{M}_R  N_L}\neq 0}$:

This is the simplest next step. However, since this involves couplings that
mix the active neutrinos $\nu_L$ and $\nu^{M}_R$ with the sterile ones,
$N_L$ and $N_R$, they will influence the mass eigenvalues of the active sector
and also their mixtures in the mass eigenstates. The knowledge one has acquired
in the determination of weak interaction couplings provides a strong
constraint on these mixing. An exhaustive phenomenological study is beyond
the scope of the present paper. However, some hints on what to expect
will be presented here.
Below, we will show two
particular examples to see the correlations between these couplings
and the mass eigenvalues and eigenstates of the active neutrino sector. 

We will concentrate on scenarios in which the heavier of the sterile Majorana
neutrinos has a mass ranging from a few MeV to a few hundreds of GeVs. 
This meant to be an example in which one can have both keV and MeV
sterile neutrinos. Let us start with
a few examples in which the heavier sterile neutrino has a mass of the order of
$M_R$.
To be definite, we shall take as in \cite{hung1} the following values
for $m_D$ and $M_R$, namely $M_R = 100\,GeV$ and $m_D = 10^{-6}\,M_R$, with
the understanding that these values are mainly for illustration purposes.

a) 
\be
\label{ex1}
\frac{M_4}{M_{R}}= \left( \begin{array}{cccc}
0&10^{-6}&0&4 \times 10^{-9} \\
10^{-6}&1&4 \times 10^{-9}&0 \\
0&4 \times 10^{-9}&0&1.8 . 10^{-4} \\
4 \times 10^{-9}&0&1.8 . 10^{-4}&1
\end{array} \right) \,.
\ee
The eigenvalues and eigenvectors are as follows (for $M_R = 100\,GeV$):
\bes
\ba
\label{leftnu}
m_1& \approx& -0.1 \,eV \, ;\\ \nonumber
\tilde{\nu}_1 &\approx& -\nu_L + 10^{-6}\,\nu^{M}_R + 2.2 \times 10^{-5}\, N_L \\ \nonumber
&&-2.2 \times 10^{-11} \, N_R \,,
\ea
\ba
\label{rightnu}
m_2& \approx& 100 \,GeV \, ;\\ \nonumber
\tilde{\nu}_2 &\approx& 10^{-6}\,\nu_L + \nu^{M}_R + 4 \times 10^{-9}\, N_L \\ \nonumber
&&-7.3 \times 10^{-13} \, N_R \,,
\ea
\ba
\label{sterile1}
m_{S1}& \approx & -3.24 \,keV \, ;\\ \nonumber
\tilde{\nu}_{S1} &\approx & -2.2 \times 10^{-5}\,\nu_L + 4 \times 10^{-9}\,\nu^{M}_R - N_L \\ \nonumber
&&1.8 \times 10^{-4} \, N_R \,,
\ea
\ba
\label{sterile2}
m_{S2}& \approx & 100 \,GeV \, ;\\ \nonumber
\tilde{\nu}_{S2} & \approx & 4 \times 10^{-9}\,\nu_L + 0\,\nu^{M}_R + 1.8 \times 10^{-4} \, N_L \\ \nonumber
&&+N_R \,,
\ea
\ees



b) We now present another example in which the heavier sterile neutrino mass
could be in the MeV range, in particular 10 MeV . For the
sake of comparison, we will show an example in which the lighter of the sterile
neutrinos has a mass around $3.25\,keV$ and $\sin \theta =2.23 \times 10^{-5}$.
\be
\label{ex2}
\frac{M_4}{M_{R}}= \left( \begin{array}{cccc}
0&10^{-6}&0&4 \times 10^{-11} \\
10^{-6}&1&4 \times 10^{-11}&0 \\
0&4 \times 10^{-11}&0&1.8 . 10^{-6} \\
4 \times 10^{-11}&0&1.8 . 10^{-6}&0.0001
\end{array} \right) \,.
\ee
with
\bes
\ba
\label{leftnu2}
m_1& \approx& -0.1 \,eV \, ;\\ \nonumber
\tilde{\nu}_1 &\approx& -\nu_L + 10^{-6}\,\nu^{M}_R + 2.2 \times 10^{-5}\, N_L \\ \nonumber
&&-3.6 \times 10^{-11} \, N_R \,,
\ea
\ba
\label{rightnu2}
m_2& \approx& 100 \,GeV \, ;\\ \nonumber
\tilde{\nu}_2 &\approx& -10^{-6}\,\nu_L - \nu^{M}_R - 4 \times 10^{-11}\, N_L \\ \nonumber
&&+2 \times 10^{-14} \, N_R \,,
\ea
\ba
\label{sterile12}
m_{S1}& \approx & -3.24 \,keV \, ;\\ \nonumber
\tilde{\nu}_{S1} &\approx & -2.2 \times 10^{-5}\,\nu_L + 16.4 \times 10^{-11}\,\nu^{M}_R - N_L \\ \nonumber
&&+0.018 \, N_R \,,
\ea
\ba
\label{sterile22}
m_{S2}& \approx & 10 \,MeV \, ;\\ \nonumber
\tilde{\nu}_{S2} &\approx & -4 \times 10^{-7}\,\nu_L + 10^{-12}\,\nu^{M}_R - 0.018 \, N_L \\ \nonumber
&&-N_R \,.
\ea
\ees

We would like to mention in passing that, as the mass of the heavier sterile neutrino increases, the mixing 
among the sterile neutrinos also increases in the mass matrix.


\ei

\item {\bf Some comments on the numerics}:

We wish to make two remarks concerning the values chosen for the matrix elements above.
The upper $2 \times 2$ bloc in (\ref{ex1}) and (\ref{ex2}) was taken from \cite{hung1}
and was chosen to give one electroweak scale mass eigenstate and one light ($\sim 0.1\,eV$)
eigenstate for the active neutrino sector. The lower $2 \times 2$ bloc for the sterile
sector was basically chosen phenomenologically to give a keV eigenstate and a heavier
one with a mass ranging from 10 MeV to 100 GeV. Basically the elements $m_{\nu_L N_R}$ and
$m_{\nu^{M}_R  N_L}$ will then determine the mixing angles between the active and
sterile sectors. They were chosen in such a way as to obey the various constraints
imposed on sterile neutrinos \cite{smirnov,kusenko}. We shall come back to
this aspect in the section on sterile neutrinos.

One might ask about the reasons why the mixing parameters between the active and sterile sectors
in the mass matrices could be so small considering the fact that $m_{\nu_L N_R}$ (\ref{nu}) and
$m_{\nu^{M}_R  N_L}$ (\ref{numir}) are proportional to the same VEVs which give masses to the 
charged leptons and quarks. One first notices that even if we set $m_{\nu_L N_R} =0$
and $m_{\nu^{M}_R  N_L} =0$ giving rise to two sets of relations, there would be {\em no}
cancellations in the expressions for the charged lepton and quark masses. As a result, one might, from a
phenomenological viewpoint set $m_{\nu_L N_R} \sim m_{\nu^{M}_R  N_L} \sim O(10^{-11}\,M_R)$ for
example. On the other hand the mixing $m_D$ (Eq. (\ref{dirmatrix})) and $m^{N}_D$ (Eq. (\ref{dirN}))
involve scalars which are $SU(2)_V$ singlets which, as it has been argued in \cite{hung1}, could
have vacuum expectation values smaller than the electroweak scale. There would be no need of cancellations
there in contrast with $m_{\nu_L N_R}$ and $m_{\nu^{M}_R  N_L}$.

\ei
A quick look at (\ref{ex1}) and (\ref{ex2})  reveals that, as the mass scales of the sterile sector
decreases, the mixing between the active and sterile sectors also decreases within
the context of our numerical examples. The astrophysical implications
of these results will be discussed below. Needless to say that these examples are shown to illustrate
some of the relationships between the sterile masses and their mixing with the active sector.

One last comment is in order here. The above numerical examples dealt with the overall mass scales
and, in that sense, would look like a one generation case. A more ``realistic'' scenario would involve
the usual three (or more) families. Nevertheless, one would expect that the above masses and mixing
would not be changed much when three or more generations are involved. One might also expect that,
if the heaviest among the light sterile neutrinos has a mass of a few keVs, some of the remaining
eigenstates might be much lighter, even having eV masses. 

\subsection{Masses of the charged mirror fermions}

We have alluded above to the fact that mirror quarks and leptons should be
heavier than their SM counterparts since they have not been observed so far.
It is fair to ask the question: Why should they be heavier than the SM
particles? First, let us remind ourselves that in this model the
SM and mirror fermions are coupled to different Higgs scalars. In
principle, there is no reason why the mass pattern of the two sectors
should be similar. 

It is without any doubt that the problem of fermion masses is one
of the biggest mysteries of the SM and, although there are many
models, no satisfactory answer has been found and widely accepted. However,
to obtain some hint on why the mirror fermions are heavier than
the SM particles, one might for instance take some ansatz that could
``fit'' the SM mass pattern and try to see how to adopt it to the
mirror sector. For simplicity, let us use the ansatz of \cite{rosner}
(see also \cite{fritzsch}).
Also for simplicity we will focus on the quark sector in this discussion.

The hierarchical ansatz of \cite{rosner} is simply the following matrix (ignoring
the phase factors)
\be
\label{rosner}
{\cal M}_H = m_3\left( \begin{array}{ccc}
0 &\epsilon^3&0 \\
\epsilon^3&\epsilon^2&\epsilon^2 \\
0&\epsilon^2&1
\end{array} \right) \,,
\ee
with the mass eigenvalues to order $\epsilon^4$ being
$-m_3\,\epsilon^4$, $m_3\,\epsilon^2$ and $m_3\,(1+\epsilon^4)$. \cite{rosner}
used $\epsilon_u=0.07$ and $\epsilon_d =0.21$ to reproduce the phenomenological mass hierarchies
at the scale $M_Z$. 

Let us assume a similar ansatz for the mirror quark sector
\be
\label{mirmass}
{\cal M}_M = m_M\left( \begin{array}{ccc}
0 &\epsilon_M^3&0 \\
\epsilon_M^3&\epsilon_M^2&\epsilon_M^2 \\
0&\epsilon_M^2&1
\end{array} \right) \,.
\ee
Since $m_M$ cannot be too different from the electroweak scale, a value
of $\epsilon_M$ similar to those of the SM quarks would engender light
mirror quarks. To avoid this, $\epsilon_M$ would have to be
significantly different from $\epsilon_u$ and $\epsilon_d$. Furthermore,
in order to satisfy constraints coming from the T parameter (to be discussed
along with the S parameter in Section (\ref{pheno})), we will assume,
for the sake of discussion, that the up and down mirror quark sectors are
``degenerate'', namely $\epsilon^{u}_M \sim \epsilon^{d}_M$. What
if the mass difference between the SM and mirror sectors is due to
the disparity between the scale $M_{LR}$ of the breakdown 
$SU(2)_L \otimes SU(2)_R \rightarrow SU(2)_V$ and $M_Z$? Let us remind ourselves
that above $M_{LR}$, the SM and mirror fermions have separate gauge interactions,
$SU(2)_L$ and $SU(2)_R$ respectively, while below that scale they interact with
the same gauge bosons of $SU(2)_V$. We make the following {\em ansatz}:
\be
\label{eps}
\epsilon^{u}_M \sim \epsilon^{d}_M \sim \frac{M_{LR}}{M_Z}\,\epsilon_{SM} \,,
\ee
where $\epsilon_{SM}$ is some value between $\epsilon_u$ and $\epsilon_d$.
We take as an example $\frac{M_{LR}}{M_Z}=10$, as discussed in the following section.
The eigenvalues corresponding to $\epsilon^{u}_M \sim 0.8,0.9$ are
(1) $m_M(-0.37,0.46,1.55)$, (2) $m_M(-0.56,0.51,1.86)$ respectively. For
$m_M \sim O(\geq 250\,GeV)$, all these mirror quarks are heavy. For example,
with $m_M=350\,GeV$ and $\epsilon^{u}_M \sim 0.9$, one obtains the following three mass
eigenvalues: $(-196,179,651)\,GeV$.
We will briefly
discuss the constraints on such masses in the section on phenomenology.
A similar consideration can be applied to the mirror lepton sector yielding heavy mirror
leptons.  

The above discussion is one of the probably many possibilities of rendering
the mirror fermions heavy. 
We now turn our attention to phenomenological constraints and implications of the model.

\section{Constraint from $\sin^{2}\theta_{W}(M_Z)$}
\label{sin2thetaw}

As we have discussed above, the SM is embedded into a PUT gauge group of the form
$G=SU(4)_{PS} \otimes SU(2)_L \otimes
SU(2)_R \otimes SU(2)^{\prime}_L \otimes SU(2)^{\prime}_R$. 
The interpretation of the fermion content in the present model is however
very different from that used for the same gauge group in \cite{put1,put2}. As
a result, the pattern of symmetry breaking and the computation of
$\sin^{2}\theta_{W}(M_Z)$ will be somewhat different here. The main
purpose for computing $\sin^{2}\theta_{W}(M_Z)$ is to constraint the
Petite Unification mass scale.

The computation of $\sin^{2}\theta_{W}(M_Z)$ in the breaking pattern
(\ref{pattern}) and, in particular, (\ref{LRbreak}), is a little more
complicated than a similar one in \cite{put1,put2}. As a result, a certain caution
is warranted. The difference with \cite{put1,put2} lies with the fact that
the weak $SU(2)$ group there was simply one of the $SU(2)$'s, namely
$SU(2)_L$. This was referred to as an ``unlocked'' case. In the
present model, the weak $SU(2)$ group comes from the breaking
$SU(2)_L \otimes SU(2)_R \rightarrow SU(2)_V$. In the language of
\cite{put1,put2}, this is a ``locked'' case and because of this,
as we shall see below, the scales $\tilde{M}$ and $M$ turn out
to be quite large. Our strategy for computing those scales will be
as follows. First, we will derive an expression for
$\sin^{2}\tilde{\theta}_{W}(M_{LR})$ which depends on the three scales
$M_{LR}$, $\tilde{M}$ and $M$. (Notice that $\tilde{\theta}_{W}$ refers
to a slightly different angle than the usual one.) We then evolve
the SM $\sin^{2}\theta_{W}$ from its experimental
value at $M_Z$ to a value at $M_{LR}$. Next, we derive a
relation between $\sin^{2}\tilde{\theta}_{W}(M_{LR})$ and
$\sin^{2}\theta_{W}(M_{LR})$. Using this relation, we then determine
the possible values for the aforementioned three scales.

Unlike \cite{put1,put2}, our basic equations will start from
the scale $M_{LR}$ instead of $M_Z$. This will be matched
with the evolution of the $SU(3)_c \otimes SU(2)_V \otimes U(1)_Y$
couplings up from $M_Z$ to $M_{LR}$. 
To be precise in our definitions, we first give a list of
notations for the gauge couplings at various mass scales.

\bi

\item  The group $SU(3)_{c}(g_3) \otimes SU(2)_V (g_2) \otimes
U(1)_{Y}(g^\prime)$ at $M_{Z}$.

\item The group $SU(3)_{c}(g_3) \otimes SU(2)_L (g_W) \otimes
SU(2)_R (g_W) \otimes U(1)_{Y}(g^\prime)$ at $M_{LR}$.

\item The group $SU(3)_{c}(g_3) \otimes SU(2)_L (g_W) \otimes
SU(2)_R (g_W) \otimes SU(2)^{\prime}_L (g_W) \otimes SU(2)^{\prime}_R (g_W) 
\otimes U(1)_{S}(\tilde{g}_S)$ at $\tilde{M}$.

\item The group $SU(4)_{PS}(g_S) \otimes SU(2)_L (g_W) \otimes
SU(2)_R (g_W) \otimes SU(2)^{\prime}_L (g_W) \otimes SU(2)^{\prime}_R (g_W)$ 
at $M$.

\ei

Following the procedure of and using the same notations as in \cite{put1,put2}, 
and making use of the definitions
of $T_{2V}$ (Eq. (\ref{su2v})), the $SU(2)_V$
gauge coupling is written at $M_{LR}$ in terms of $g_W$ as
\be
\label{g2}
\frac{1}{g_{2}^{2}(M_{LR}^{2})}=\frac{2}{g_{W}^{2}(M_{LR}^{2})}\,,
\ee 
where the factor of $2$ in (\ref{g2}) comes from (\ref{su2v}). This will be used
below to obtain a match at $M_{LR}$. Since we have
$SU(2)^{\prime}_L (g_W) \otimes SU(2)^{\prime}_R (g_W) 
\otimes U(1)_{S}(\tilde{g}_S) \rightarrow U(1)_Y(g^{\prime})$ at $\tilde{M}$, one has
\be
\label{gp}
\frac{1}{(g^{\prime})^{2}(\tilde{M}^{2})}=\frac{2}{g_{W}^{2}(\tilde{M}^{2})} +
\frac{C_{S}^2}{\tilde{g}_{S}^{2}(\tilde{M}^{2})}\,,
\ee
where we have used Eq. (\ref{Y/2p}) and where $C_{S}^2 = 2/3$. Furthermore, at the
scale $M$, one has
\be
\label{puteq}
g_{3}(M^{2})= \tilde{g}_{S}(M^{2})= g_{S}(M^{2}) \,.
\ee
 
One can now examine the evolution of the gauge couplings from $M_{LR}$ to $\tilde{M}$ in details.

Since the particle content of the $SU(2)$'s groups are symmetric and
since it is assumed that the fermions and scalars have masses less than
$M_{LR}$, one can use either $SU(2)_L$ or $SU(2)_R$ to study the evolution of the couplings from
$M_{LR}$ to $\tilde{M}$. The basic equations used here are
\be
\label{g2p}
\frac{1}{g_{W}^{2}(M_{LR}^{2})}=\frac{(C_W^{\prime})^2}{g_{W}^{2}(\tilde{M}^{2})} + 2\, b_2 \, \ln(\frac{\tilde{M}}{M_{LR}})\,,
\ee 
\be
\label{gprime}
\frac{1}{(g^{\prime})^{2}(M_{LR}^{2})}= \frac{C_W^2 }{g_{W}^{2}(\tilde{M}^{2})} +
\frac{C_{S}^2}{\tilde{g}_{S}^{2}(\tilde{M}^{2})} + 2\, b_1 \, \ln(\frac{\tilde{M}}{M_{LR}})\,,
\ee
\be
\label{g3}
\frac{1}{g_{3}^{2}(M_{LR}^{2})}=\frac{1}{g_{S}^{2}(M^{2})} + 2\, b_3 \, \ln(\frac{M}{M_{LR}})\,,
\ee 
\be
\label{gS}
\frac{1}{\tilde{g}_{S}^{2}(\tilde{M}^{2})}=\frac{1}{g_{S}^{2}(M^{2})} + 2\, \tilde{b} \, \ln(\frac{M}{\tilde{M}})\,,
\ee 
where $C_{S}^2 = 2/3$, $(C_W^{\prime})^2=1$, $C_W^2=2$ and where
\be
\label{b1}
b_1 = \frac{1}{48 \pi^2}(\frac{20}{3}n_G +7) \, ,
\ee
\be
\label{b2}
b_2 = \frac{1}{48 \pi^2}(2\,n_G +5 - 22) \, ,
\ee
\be
\label{b3}
b_3 = \frac{1}{48 \pi^2}(4\,n_G  - 33) \, ,
\ee
\be
\label{btilde}
\tilde{b} = \frac{1}{48 \pi^2}(4\,n_G) \, ,
\ee
with $n_G$ (= left-handed plus right-handed)  
is the number of $SU(2)_V$ doublets. (The number of families is $n_G/2$.) In
addition we also list the coefficient related to $SU(2)_V$ for the use
in the evolution of the $SU(2)_V$ coupling from $M_Z$ to $M_{LR}$:
\be
\label{b2v}
b_{2V} = \frac{1}{48 \pi^2}(4\,n_G +7 - 22) \, ,
\ee

A few words concerning the different factors in (\ref{b1}, \ref{b2}, \ref{b3}, \ref{btilde},
\ref{b2v}) are in order. 
First, $n_G$ refers to the number of $SU(2)_V$ doublets and that includes
{\em both} left-handed and right-handed fermions. Now, $b_2$ refers to $SU(2)_L$ (or $SU(2)_R$)
and, as a result, $n_{L,R} = n_G/2$ resulting in the factor $2\,n_G$ in (\ref{b2})
while it is $4n_G$ in (\ref{b2v}).
Second, the factors $7$ in (\ref{b1}) and $5$ in (\ref{b2}) comes from the counting of the
number of scalar degrees of freedom as follows. From $\Phi_S$, $\Phi^{M}_S$ and the $SU(3)_c$-singlet
parts of $\Phi^{A}$ and $\Phi^{A}_{M}$, one obtains $8$ Higgs doublets which contribute 
a factor of $4$ to $b_1$. In addition, the two Higgs triplets contribute a factor $1+2=3$
to $b_1$ giving a total of $7$ from the scalar sector. 
On the other hand, above $M_{LR}$, only $4$ Higgs doublets contribute to $b_2$ since
one now has e.g. $SU(2)_L$ instead of $SU(2)_V$ while below $M_{LR}$ all $4$ Higgs
doublets contribute to $b_{2V}$.

Eqs. (\ref{b1},\ref{b2},\ref{b3},\ref{btilde}) are similar in forms to the ones used in \cite{put1}
to derive $\sin^{2}\theta_{W}(M_{Z}^2)$ except that now we will use them to derive an expression
for $\sin^{2}\tilde{\theta}_{W}(M_{LR}^2)$ and a relationship between these two
quantities. Although the $U(1)_{em}$ gauge coupling is transmogrified into
the $SU(2)_{V} \otimes U(1)_Y$ gauge couplings above the electroweak scale (or
$M_Z$), we will keep the same notation above that scale, namely
\be
\label{em1}
\frac{1}{e^{2}(M_{LR}^{2})}= \frac{1}{g_{2}^2(M_{LR}^{2})} + \frac{1}{(g^{\prime})^{2}(M_{LR}^{2})} \,.
\ee
Let us define a similar quantity involving $g_W$, namely
\be
\label{em2}
\frac{1}{\tilde{e}^{2}(M_{LR}^{2})}= \frac{1}{g_{W}^2(M_{LR}^{2})} + \frac{1}{(g^{\prime})^{2}(M_{LR}^{2})} \,,
\ee
and
\be
\label{alpha2}
\tilde{\alpha} (M_{LR}^{2}) \equiv \frac{\tilde{e}^{2}(M_{LR}^{2})}{4\,\pi}\,.
\ee
Let us recall the definition of $\sin^{2}\theta_{W}(M_{Z}^2)$:
\be
\label{sintheta1}
\sin^{2}\theta_{W}(M_{Z}^2) \equiv \frac{e^{2}(M_{Z}^{2})}{g_{2}^2(M_{Z}^{2})} \,,
\ee
with a similar expression evaluated at $M_{LR}$. Let us now define
$\sin^{2}\tilde{\theta}_{W}(M_{LR}^2)$ as
\be
\label{sintheta2}
\sin^{2}\tilde{\theta}_{W}(M_{LR}^2) \equiv \frac{\tilde{e}^{2}(M_{LR}^{2})}{g_{W}^2(M_{LR}^{2})} \,.
\ee
Obviously $\sin^{2}\tilde{\theta}_{W}(M_{LR}^2)$ is not the same as
$\sin^{2}\theta_{W}(M_{Z}^2)$. One can easily derive a relation between the two
as follows
\be
\label{sintheta2p}
\sin^{2}\tilde{\theta}_{W}(M_{LR}^2) = \frac{\sin^{2}\theta_{W}(M_{LR}^2)/2}
{1-(\sin^{2}\theta_{W}(M_{LR}^2)/2)}\,,
\ee
upon using Eqs. (\ref{em1},\ref{em2},\ref{sintheta1},\ref{sintheta2}). Since we will be comparing the
predicted results with the usual $\sin^{2}\theta_{W}(M_{Z}^2)$, one can rewrite
(\ref{sintheta2p}) as
\be
\label{sintheta1p}
\sin^{2}\theta_{W}(M_{LR}^2) = \frac{2\,\sin^{2}\tilde{\theta}_{W}(M_{LR}^2)}
{1+\sin^{2}\tilde{\theta}_{W}(M_{LR}^2)}\,.
\ee

From Eqs. (\ref{g2p},\ref{gprime},\ref{g3},\ref{gS},\ref{em2},\ref{alpha2}), one can readily derive the following formula for
$sin^{2}\tilde{\theta}_{W}(M_{LR}^2)$:
\ba
\label{basic}
\sin^{2}\tilde{\theta}_{W}(M_{LR}^2)& =& \sin^{2}\tilde{\theta}_{W}^{0}\{
1-C_{S}^2 \frac{\tilde{\alpha} (M_{LR}^{2})}{\alpha_{S}(M_{LR}^{2})} \nonumber \\
&&-8\pi \tilde{\alpha} (M_{LR}^{2})[K\, \ln(\frac{\tilde{M}}{M_{LR}}) + K^{\prime}
\ln(\frac{M}{\tilde{M}})]\}\,, \nonumber \\
\ea
where, in the parlance of \cite{put1},
\be
\label{sintheta0}
\sin^{2}\tilde{\theta}_{W}^{0} = \frac{(C_W^{\prime})^2}{C_W^2 + (C_W^{\prime})^2} = \frac{1}{2+1} 
=\frac{1}{3} \,,
\ee
\be
\label{K}
K=b_1-2\,b_2-\frac{2}{3}b_3 =\frac{125}{96\,\pi^2}\,,
\ee
\be
\label{K'}
K^{\prime} =C_{S}^2\,(\tilde{b}-b_3)=\frac{22}{48\,\pi^2} \,,
\ee
and where we have used Eqs. (\ref{b1},\ref{b2},\ref{b3},\ref{btilde}). Notice the interesting fact that the
dependence on $n_G$ drops out in both $K$ and $K^{\prime}$.

To proceed with (\ref{basic}), we need to evaluate $\tilde{\alpha} (M_{LR}^{2})$ and
$\alpha_{S}(M_{LR}^{2})$. From the above equations, one can relate $\tilde{\alpha} (M_{LR}^{2})$
to the following measured quantities at $M_Z$ as follows
\ba
\label{alphatil2}
\tilde{\alpha}^{-1} (M_{LR}^{2})&=& \alpha^{-1}(M_Z^2)(1-\frac{1}{2}
\sin^{2}\theta_{W}(M_{Z}^2))  \nonumber \\ 
&&- 8\,\pi(b_1+\frac{1}{2}b_{2V})\ln(\frac{M_{LR}}{M_Z}) \,. 
\ea
In (\ref{alphatil2}), we will use $\alpha^{-1}(M_Z^2) = 127.934$,
$\sin^{2}\theta_{W}(M_{Z}^2) = 0.23113$. 
Furthermore, with the assumption of three or four (SM and mirror) families
i.e. $n_G = 6,8$, it can easily be seen that
$\alpha_{S}(M_{LR}^{2}) \approx \alpha_{S}(M_{Z}^{2}) \approx 0.117$. This last point is interesting
on its own: In our model, QCD is nearly scale-invariant above $M_Z$!

The next step involves the extraction from $\sin^{2}\tilde{\theta}_{W}(M_{LR}^2)$ of the
values of $\sin^{2}\theta_{W}(M_{Z}^2)$. Several steps are involved in this computation.
\bi

\item From $\sin^{2}\tilde{\theta}_{W}(M_{LR}^2)$, we can extract 
$\sin^{2}\theta_{W}(M_{LR}^2)$ via Eq. (\ref{sintheta1p}).

\item Next we calculate $\alpha^{-1} (M_{LR}^{2})$ using
\ba
\label{al1}
\alpha^{-1} (M_{LR}^{2})& =& \alpha^{-1} (M_{Z}^{2})-8\,\pi\,(b_1+b_{2V})\,
\ln(\frac{M_{LR}}{M_Z}) \nonumber \\
&=& \alpha^{-1} (M_{Z}^{2}) -\frac{1}{6\,\pi}(\frac{32}{3}\,n_G-8)\,
\ln(\frac{M_{LR}}{M_Z}) \,. \nonumber \\
\ea

\item $\alpha^{-1}_{2} (M_{LR}^{2})$ is obtained from
\be
\label{al2lr}
\alpha^{-1}_{2} (M_{LR}^{2}) =\sin^{2}\theta_{W}(M_{LR}^2)\,
\alpha^{-1} (M_{LR}^{2}) \,.
\ee 

\item Next we compute
\ba
\label{al2lr2}
\alpha^{-1}_{2} (M_{Z}^{2})&=&\alpha^{-1}_{2} (M_{LR}^{2})+
8\pi\,b_{2V}\,\ln(\frac{M_{LR}}{M_Z}) \nonumber \\
&=&\alpha^{-1}_{2} (M_{LR}^{2})+
\frac{1}{6\,\pi}(4\,n_G-15)\,
\ln(\frac{M_{LR}}{M_Z}) \,.\nonumber \\
\ea

\item Finally, we obtain
\be
\label{sinfin}
\sin^{2}\theta_{W}(M_{Z}^2) = \alpha^{-1}_{2} (M_{Z}^{2})\,\alpha (M_{Z}^{2})\,.
\ee

\ei

As an example, we will show
two cases corresponding to two values of the ratio $\frac{M_{LR}}{M_Z}$. These are
shown in the tables below.
\begin{table}
\label{tab1_8}
\caption{Average values of $\tilde{M}$ and $M$ for $\frac{M_{LR}}{M_Z} =10$ and $n_G=8$ subjected to
the constraint $0.2308 \leq \sin^{2}\theta_{W}(M_{Z}^2) \leq 0.2314 $} 
\begin{ruledtabular}
\begin{tabular}{ccccc} 
$\tilde{M}$(GeV)&$9.51 \times 10^{6}$&$2.12 \times 10^{7}$&$4.75 \times 10^{7}$&$1.06 \times 10^{8}$ \\ \hline
$M$(GeV)&$9.51 \times 10^{17}$&$2.12 \times 10^{16}$&$4.75 \times 10^{15}$&$1.06 \times 10^{15}$ \\
\end{tabular}
\end{ruledtabular}
\end{table}
\begin{table}
\label{tab1_6}
\caption{Average values of $\tilde{M}$ and $M$ for $\frac{M_{LR}}{M_Z} =10$ and $n_G=6$ subjected to
the constraint $0.2308 \leq \sin^{2}\theta_{W}(M_{Z}^2) \leq 0.2314 $} 
\begin{ruledtabular}
\begin{tabular}{ccccc} 
$\tilde{M}$(GeV)&$1.16 \times 10^{7}$&$2.59 \times 10^{7}$&$5.78 \times 10^{7}$&$1.59 \times 10^{8}$ \\ \hline
$M$(GeV)&$1.16 \times 10^{17}$&$2.59 \times 10^{16}$&$5.78 \times 10^{15}$&$1.59 \times 10^{15}$ \\
\end{tabular}
\end{ruledtabular}
\end{table}
\begin{table}
\label{tab2_8}
\caption{Average values of $\tilde{M}$ and $M$ for $\frac{M_{LR}}{M_Z} =5$ and $n_G=8$ subjected to
the constraint $0.2308 \leq \sin^{2}\theta_{W}(M_{Z}^2) \leq 0.2314 $} 
\begin{ruledtabular}
\begin{tabular}{ccccc} 
$\tilde{M}$(GeV)&$7.68 \times 10^{6}$&$1.72 \times 10^{7}$&$3.83 \times 10^{7}$&$8.57 \times 10^{7}$ \\ \hline
$M$(GeV)&$7.68 \times 10^{16}$&$1.72 \times 10^{16}$&$3.83 \times 10^{15}$&$8.57 \times 10^{14}$ \\
\end{tabular}
\end{ruledtabular}
\end{table}
\begin{table}
\label{tab2_6}
\caption{Average values of $\tilde{M}$ and $M$ for $\frac{M_{LR}}{M_Z} =5$ and $n_G=6$ subjected to
the constraint $0.2308 \leq \sin^{2}\theta_{W}(M_{Z}^2) \leq 0.2314 $} 
\begin{ruledtabular}
\begin{tabular}{ccccc} 
$\tilde{M}$(GeV)&$8.79 \times 10^{6}$&$1.96 \times 10^{7}$&$4.39 \times 10^{7}$&$9.83 \times 10^{7}$ \\ \hline
$M$(GeV)&$8.79 \times 10^{16}$&$1.96 \times 10^{16}$&$4.39 \times 10^{15}$&$9.83 \times 10^{14}$ \\
\end{tabular}
\end{ruledtabular}
\end{table}
The above two examples show the range of mass scales required to give values for 
$\sin^{2}\theta_{W}(M_{Z}^2)$ consistent with experiment.

Although the above examples are far from being exhaustive, the mass scale pattern seems
to be one in which $M_{LR} \alt 1\,TeV$ ($\sim$ masses of heavy $W_{H}$), 
$\tilde{M} \sim 10^{7}\,GeV- 10^{8}\,GeV$ ($\sim$ masses of $SU(2)^{\prime}_L \otimes SU(2)^{\prime}_R$
gauge bosons), and $M\sim 10^{15}\,GeV-10^{17}\,GeV$ (the Pati-Salam unification scale). Let
us notice however that this is not a prediction for the mass scales. It is
simply an illustration of the relationships between the various scales constrained
by the experimental values of $\sin^{2}\theta_{W}(M_{Z}^2)$. It goes without saying that
$M_{LR}$ could be larger than $1\,TeV$ in which case the Pati-Salam unification scale
could be higher than $10^{17}\,GeV$.
The phenomenological implications of these values will be discussed in the next section. 

\section{Phenomenological implications}
\label{pheno}

In this section we will present a very brief discussion of the phenomenology, the
details of which will be presented elsewhere. Three main issues among several others
include the presence of mirror fermions \cite{hung1}, sterile neutrinos (in addition
to the electroweak scale right-handed neutrinos), and
heavy W-like gauge bosons.
\bi

\item {\bf Mirror fermions}:

What are the effects of the mirror fermions or,
in general, of extra chiral families on the S and T parameters? This question
has been previously studied in the literature. For example, it was found that
there are regions of parameter space in
a two-Higgs doublet model that can accommodate even {\em three additional chiral families}
\cite{he}. With the presence of Higgs triplets, it was also found that
one can have large negative contributions to S, offsetting 
possible positive contributions to S coming from extra fermion families
\cite{hunge6, hungbran}. Our model contains
eight Higgs doublets, one complex Higgs triplet and one real
Higgs triplet where the counting include only Higgs fields
that develop VEVs ((\ref{vevsm})-(\ref{vev15}) and (\ref{vevphi10})). 
Needless to say, we have the necessary ingredients to
accommodate the extra mirror families. One can even work with a four family scenario
(SM plus mirror) since one now has enough Higgs representations to offset
any non-degeneracy of the extra families. In summary, one can have
the correct S and T parameters in our model.
We will now focus on various phenomenological aspects of mirror fermions.

The crux of the model presented in \cite{hung1} is the existence of electroweak
scale right-handed neutrinos which are non-sterile. Because of this fact, one
could directly produce those right-handed neutrinos at colliders such
as the upcoming LHC or the proposed ILC and check the validity of the
see-saw mechanism. As mentioned in \cite{hung1}, one of most important
signals of the model is the presence of like-sign dileptons at colliders. Such a
signal would constitute a high-energy equivalent of neutrinoless double beta
decay.

The like-sign dilepton events can come from the following subprocesses.

1) Production and subsequent decays of electroweak scale
right-handed neutrinos:
\ba
\label{qq}
q + \bar{q} \rightarrow Z &\rightarrow& \nu^{M}_R + \nu^{M}_R \nonumber \\ 
&\rightarrow&
e^{M,\mp}_R +e^{M,\mp}_R +W^{\pm} + W^{\pm} \nonumber \\
&\rightarrow& e^{\mp}_L +e^{\mp}_L +W^{\pm} + W^{\pm}
+ \phi_S + \phi_S \,,\nonumber \\
\ea
where $e^{M,\mp}_R$ could be real or virtual depending on the
mass differences with $\nu_R$'s. The production cross section
is estimated to be $\sigma \sim 400\,fb$ for $M_R \sim 100\,GeV$ at
the LHC.
One can also have e.g.
\ba
\label{ud}
u + \bar{d} \rightarrow W^+
&\rightarrow&  \nu^{M}_R + e^{M,+}_R \nonumber \\ 
&\rightarrow& e^{M,+}_R +e^{M,+}_R +W^{-}    \nonumber \\ 
&\rightarrow& e^{+}_L +e^{+}_L +W^{-}
+ \phi_S + \phi_S \,, \nonumber \\ 
\ea
where $e$ and $e^{M}$ are generic notations for (SM and mirror) charged leptons.
In the above processes, $\phi_S$ is the singlet scalar field which would be
considered as missing energy. The W's could transform into jets or pairs of leptons.
Depending on how heavy $\phi_S$ is, the signal could be quite interesting.
If the mirror charged lepton is sufficiently long-lived (e.g. its decay 
could occur a few centimeters away from the beam pipe), the search
for like-sign dileptons with displaced vertices would constitute
perhaps a ``clean'' signal.

2) Direct production of like-sign mirror dileptons followed by like-sign
SM leptons:
\ba
\label{ww}
W^{+} + W^{+} \rightarrow  \chi^{++} &\rightarrow&  e^{M,+}_R + e^{M,+}_R \nonumber \\
&\rightarrow& e^{+}_L + e^{+}_L + \phi_S + \phi_S  \,,\nonumber \\
\ea
and similarly for the opposite sign process. Here one expects, at the LHC, a
production cross section $\sigma \sim 3\,pb$ for $M_{\chi^{++}} \sim 200\,GeV$.

In a scenario in which $\nu^{M}_R$s are all lighter than the mirror
charged leptons, the direct production
of lighter $\nu^{M}_R$s is followed by the decay into $\nu_L$s plus $\phi_S$, all of which 
will become missing energy:
\ba
\label{qq2}
q + \bar{q} \rightarrow Z &\rightarrow& \nu^{M}_R + \nu^{M}_R \nonumber \\ 
&\rightarrow& \nu_L + \nu_L + + \phi_S + \phi_S \,. \nonumber \\ 
\ea

All of the above
processes can occur and it will be interesting to disentangle the two
like-sign dilepton mechanisms. Beside the difference in cross sections,
the process (\ref{qq}, \ref{ud}) has like-sign dileptons plus e.g. one or two jets and missing
energy while the process (\ref{ww}) has a like-sign dilepton plus missing energy. This phenomenology
will be presented elsewhere.

3) Last but not least, it would be interesting to study the phenomenology associated with
mirror quarks. These quarks would be produced in colliders just like a typical SM heavy
quarks e.g. the top quark.
For the heavy mirror quarks to materialize into SM particles, one can look at the
decay process $q^{M} \rightarrow q + \phi_S$. Here one should perhaps concentrate
on displaced vertices depending on how long-lived the mirror quarks are \cite{physrep}.

\item {\bf Sterile neutrinos $N_L$ and $N_R$}:

As we have extensively discussed in Section (\ref{sterile}) various
aspects of the sterile neutrinos  $N_L$ and $N_R$, we would like to
briefly discuss what these particles could do in astrophysics and
cosmology, among others. A comprehensive study of various constraints
on the sterile neutrino sector can be found in \cite{smirnov}.

We have seen above that there are two types of sterile neutrinos in our
model: $N_L$ and $N_R$. In the two numerical examples given in Section
(\ref{sterile}), one can have a situation in which the keV sterile neutrino
is almost purely $N_L$. (Again, for simplicity, we discuss the one specie
case although the keV state could refer to the heavier among the light sterile
neutrinos.) The first question to ask here is the following: What
can a keV sterile neutrino do? The various possibilities have been extensively
discussed (for a review and a list of references, see \cite{kusenko}) and we will 
just summarize some of the relevant
points here. For definiteness, we will take as an example a 3.24 keV
sterile neutrino (mostly $N_L$) with a mixing angle to the SM light neutrino sector
of the order $2.2 \times 10^{-5}$ as shown in (\ref{ex1}) and (\ref{ex2}). 

First, the possibility of keV sterile neutrinos being
candidates for Warm Dark Matter (WDM) has been an exciting 
and active avenue of research \cite{WDM}. It is believed
that the $\Lambda$CDM scenario while extremely successful
in explaining a fair number of cosmological phenomena, appears
to have problems with structure formation, in particular
concerning the prediction of the number of dwarf galaxies which
seem to be much larger than the ``observed'' number of such
objects. The addition of WDM in the form of keV sterile
neutrinos appears to alleviate this problem. However, it has
been argued that the sterile neutrinos cannot constitute
the entire content of dark matter because of conflicts between
constraints coming from direct x-ray observations and indirect
ones coming from structure formation in the form of the so-called
Lyman-$\alpha$ forest measurements \cite{silk}. This conflict
arises within a specific production mechanism for the sterile neutrinos:
the Dodelson-Widrow (DW) mechanism \cite{dodelson} where sterile neutrinos are produced
through non-resonant oscillations with the active neutrinos.
There it was found that
for a sub-dominant fraction of sterile neutrinos $f_S = 
\frac{\Omega_s}{\Omega_{dm}}$, say $f_S=0.2$, there is a range
of allowed masses and mixing which satisfies both constraints,
namely $m_S \sim 2.5-16\, keV$ and $\sin^{2} 2\theta \sim
9 \times 10^{-11}-2.5 \times 10^{-9}$. Furthermore the
allowed region shrinks to zero at the $3\sigma$ level as
$f_S$ tends to unity. Ref.\cite{kusenko2} has reanalyzed
the the proposal that keV sterile neutrinos could be
the sources of pulsar kicks \cite{kusenko} in light of
the new constraints of \cite{silk}. Sterile neutrinos situated in the 
above range satisfy both x-ray and Ly$\alpha$ constraints and provide
a possible source for pulsar kicks. As stressed in \cite{silk},
the constraints on $f_S$ might be relaxed if mechanisms other
than the DW non-resonant oscillation for the sterile neutrino
production are invoked. For a recent discussion, see
Ref.\cite{petraki}. The lighter sterile state $N_L$
in our model fits the above discussion.

Unlike many models where the sterile neutrinos are right-handed
electroweak singlet neutrinos which participate in the see-saw
mechanism with the active left-handed neutrinos, the
sterile neutrinos in our model do not mix much with the
active ones ({\em both} left and right-handed states). 
Furthermore they come with both {\em helicities}: $N_L$ and $N_R$.
We have mentioned above the possibility that the lighter
sterile neutrino $N_L$ could have a mass of O(keV) with
the astrophysical implications that they might have. What could we say
about the heavier $N_R$? 
An extensive study of combined laboratory, astrophysical and cosmological
constraints can be found in Ref.\cite{smirnov}. An earlier bound based
on accelerator and super-Kamiokande constraints can be found
in \cite{kusenko3}.
For simplicity, we will just
mention the two examples discussed in Section (\ref{sterile}).
The first example has $m_{N_R}=100\,GeV$ and the mixing
with with the SM $\nu_L$ being $4 \times 10^{-9}$. The
second example has $m_{N_R}=10\,MeV$ and a mixing with
the SM $\nu_L$ being $4 \times 10^{-7}$. For the first example,
one has $\sin^{2}\theta_S \approx 1.6 \times 10^{-17}$. From 
\cite{smirnov}, one can see that this is well inside the allowed regions
for $N_R \leftrightarrow \nu_e$, $N_R \leftrightarrow \nu_{\mu}$
and $N_R \leftrightarrow \nu_{\tau}$. For the second example
with $m_{N_R}=10\,MeV$, one has $\sin^{2}\theta_S \approx 1.6 \times 10^{-13}$.
This is well inside the region forbidden by the CMB data \cite{smirnov} although
it is allowed by accelerator and super-Kamiokande data \cite{kusenko3}.
For the kind of mixing angles considered in our examples,
$m_{N_R}$ appears to be bounded from below by a few hundreds of MeVs.
In low reheating cosmological scenarios, it is claimed that
for $m_S > 30\, MeV$, cosmological bounds no longer apply
\cite{gelmini}. In our rather simple analysis here, the mixing
of $N_R$ with the active sector is quite small and it is not
clear how one could detect such an object.

\item {\bf Heavy W-like gauge bosons}:

Unlike extended models in which the electroweak $SU(2)$ group is
simply $SU(2)_L$ with other gauge groups being spontaneously broken
at a larger scale than the electroweak one, our model is
rather different in that the SM $SU(2)$ group comes from
$SU(2)_L \otimes SU(2)_R \rightarrow SU(2)_V$. At the scale
$M_{LR}$, one has the following interaction Lagrangian
\ba
\label{lag1}
{\cal L}_W &=& g_W \vec{J}^{\mu}_L\,\cdot \vec{W}_{L\mu} +
g_W \vec{J}^{\mu}_R\,\cdot \vec{W}_{R\mu} \nonumber \\
&=& \frac{g_W}{\sqrt{2}}(\vec{J}^{\mu}_L + \vec{J}^{\mu}_R)
\cdot \frac{\vec{W}_{L\mu} + \vec{W}_{R\mu}}{\sqrt{2}} \nonumber \\
&&+ \frac{g_W}{\sqrt{2}}(\vec{J}^{\mu}_L - \vec{J}^{\mu}_R)
\cdot \frac{\vec{W}_{L\mu} - \vec{W}_{R\mu}}{\sqrt{2}} \,,
\ea
where
\be
\label{JL}
\vec{J}^{\mu}_L = \bar{f}\gamma^{\mu}\frac{(1-\gamma_5)}{2}\vec{T}_L \, f\,,
\ee
and
\be
\label{JR}
\vec{J}^{\mu}_R = \bar{f}^{M}\gamma^{\mu}\frac{(1+\gamma_5)}{2}\vec{T}_R \, f^{M}\,.
\ee
 In (\ref{JL}) and (\ref{JR}), $f$ and $f^{M}$ refer to SM and mirror fermions
respectively. From (\ref{lag1}), one can identify
\be
\label{W}
\vec{W}_{V\mu}= \frac{\vec{W}_{L\mu} + \vec{W}_{R\mu}}{\sqrt{2}}\,,
\ee
as the electroweak gauge bosons while the orthogonal combination
\be
\label{Wprime}
\vec{W}^{\prime}_{\mu}= \frac{\vec{W}_{L\mu} - \vec{W}_{R\mu}}{\sqrt{2}}\,,
\ee
represents the heavy $W^{\prime}$ with mass $\sim M_{LR}$. Notice that
the electroweak gauge coupling denoted by the usual $g_2$ is related to
$g_W$ by
\be
\label{g2pprime}
g_2 = \frac{g_W}{\sqrt{2}}\,,
\ee
which is the same as Eq. (\ref{g2}). Below $M_{LR}$, one can write the
interaction Lagrangian involving $W$ and $W^{\prime}$ as
\be
\label{wpint}
{\cal L}_{W} =g_2\,(\vec{J}^{\mu}_L + \vec{J}^{\mu}_R)\cdot \vec{W}_{V\mu}+
g_2\,(\vec{J}^{\mu}_L - \vec{J}^{\mu}_R)\cdot 
\vec{W}^{\prime}_{\mu} \,.
\ee
The SM and mirror fermions interact with $W^{\prime}$ with the same strength
but with a different sign. One can perhaps exploit this sign difference
to isolate the contribution from $W^{\prime}$.
From (\ref{wpint}), one can see that, at low energies,
the ``fermi'' constant involving $W^{\prime}$ is related to the usual Fermi
constant by
\be
\label{fermi}
\frac{G^{\prime}}{\sqrt{2}} = \frac{G_F}{\sqrt{2}}\,(\frac{M_{W}^2}{M_{W^{\prime}}^2})\,.
\ee
Looking back at Section (\ref{sin2thetaw}), one can deduce that 
$\frac{G^{\prime}}{\sqrt{2}} \sim 10^{-2}\frac{G_F}{\sqrt{2}}$.

The present bounds on $W^{\prime}$ with standard couplings to the SM fermions
are $m> 800\,GeV$ ($95\%$ CL) for $W \rightarrow e\nu,\mu \nu$ and
$m> 825\,GeV$ ($95\%$ CL) for $Z^{\prime}$ \cite{PDG}, although the latter's
bound is more model dependent. The phenomenology of these gauge bosons
will be presented elsewhere.

\item {\bf Other consequences}:

First, we notice that the previous consideration of the group
$SU(4)_{PS} \otimes SU(2)_L \otimes
SU(2)_R \otimes SU(2)^{\prime}_L \otimes SU(2)^{\prime}_R$ where
the PUT scale was of the order of TeVs leading to severe a violation
of the upper bound on $K_L \rightarrow \mu e$ by several orders
of magnitude \cite{put2}. This is no longer the case in the present scenario
since the scale of the spontaneous breakdown of $SU(4)_{PS} \rightarrow
SU(3)_c \otimes U(1)_S$, $M$, is of the order of a typical GUT scale (\ref{sin2thetaw}).
In fact flavor changing neutral current (FCNC) processes involving
$SU(4)_{PS}/SU(3)_c \otimes U(1)_S$ gauge bosons are totally suppressed
in our present scenario.

Second, notice that the Higgs field which participates in the Majorana
mass term for the (active) right-handed neutrino is
$\Phi_{10} = (\bar{10}= 1 + \bar{3} + \bar{6}, 1, 3, 1, 1)$ (\ref{10}). One
expects the color non-singlet parts of $\Phi_{10}$ ($\bar{3}$ and $\bar{6}$) 
acquire a large mass of the order of $M \sim 10^{15}-10^{17}\,GeV$. Although the 
$SU(4)_{PS}/SU(3)_c \otimes U(1)_S$ gauge bosons do not induce proton decay,
it can occur through the exchange of these color non-singlet scalars. This study
will be presented elsewhere but one can briefly summarize the situation here
by stating that modes such as $p \rightarrow \pi^{0} e^{+}$,
$p \rightarrow \pi^{+} \bar{\nu}$ and $p \rightarrow K^{+} \bar{\nu}$ can in
principle all occur and the decay rate can be under control.

\ei 

\section{Families from spinors}
\label{spinor}

\bi

\item {\bf Possibility of a fourth generation}:
In Section (\ref{sin2thetaw}), we have discussed the computation
of $\sin^{2}\theta_{W}(M_Z)$ for three and four generations. One might
be perhaps a little puzzled about the reason for even discussing the
four generation case. And there is always the quintessential question
of why the fourth active neutrino has to be much more massive than the
other three (at least half the Z mass).
Below we will present some phenomenological
and theoretical reasons for why one might seriously consider the 4th family.

There is a quintessential question of why
there exists three families of quarks and leptons. Whether or not
there are more than three generations, in particular a fourth family,
is another question that has been entertained over the years \cite{fourth}
but whose possibility was generally dismissed because of an apparent
conflict with electroweak precision data, notwithstanding the fact
that none has been observed so far. However, recent studies have
revealed that not only was a fourth generation not ruled out by
precision data but it might even have implications concerning
the SM Higgs boson mass \cite{kribs} and perhaps
rare B decays \cite{hou}. Its existence might even help bringing
in coupling constant unification at the two-loop level \cite{hunggut}. 
Furthermore, another recent analysis
of experimental constraints on a fourth generation of quarks
presented regions of allowed masses and mixing angles
(between the fourth and the other three generations) which are
more flexible than the widely quoted mass lower bounds \cite{pqsher}. 
If the fourth family is not excluded experimentally and might
even be detected in the future, one is again faced with the
puzzle of family replication. Are there guiding principles 
which might help us to partially unravel this mystery?  

Let us first notice that a spinor of $SO(2n + 2m)$ decomposes
into $2^{m-1} \psi_{+} + 2^{m-1} \psi_{-}$ of the subgroup
$SO(2n)$, where $\psi_{+,-}$ are two distinct spinors of
$SO(2n)$. This fact has been exploited in a number of papers
on family replication \cite{famrep,goran2}. We will present a heuristic
argument why, in our framework, it is desirable to have four generations. 

One has $SO(4) \approx SU(2) \otimes SU(2)$ at the Lie algebra level.
(Group-theoretically, one actually has $SO(4) \approx (SU(2) \otimes SU(2))/Z_2$.)
Let $\psi_{+} = (2,1)$ and $\psi_{+} = (1,2)$ of $SU(2) \otimes SU(2)$
and let this represent one family.
In consequence, a spinor of $SO(2m+4)$ decomposes into
$2^{m-1} \psi_{+} + 2^{m-1} \psi_{-}$ of $SO(4)$ or $2^{m-1}$ families. If, 
in addition, one
requires that $SO(2m+4)$, as a gauge theory, should be anomaly-free, one
observes the following features: (1) $m=1$ corresponds to one family and
$SO(6)$ which is {\em not} anomaly-free; (2) $m=2$ corresponds to two families;
(3) $m=3$ corresponds to {\em four} families, etc...Case (1) is ruled out by
the anomaly-freedom requirement; Case (2) is ruled out by {\em observation}.
This leaves us with the simplest allowed case of $m=3$ which corresponds to
{\em four} families. This would correspond to the group $SO(10)$
Notice that in this case one would have $SO(10) \rightarrow SU(4)_H
\otimes SU(2) \otimes SU(2)$ where the subscript $H$ denotes ``horizontal''
or ``family''.
One might envision the following group:
\ba
\label{family}
&&SU(4)_{PS} \otimes SO(10) \otimes SO(10)^{\prime}
\rightarrow \nonumber \\
&&SU(4)_{PS} \otimes SU(4)_H \otimes SU(2)_L \otimes SU(2)_R \nonumber \\
&&\otimes SU(4)_H^{\prime} \otimes SU(2)_L^{\prime} \otimes SU(2)_R^{\prime} \,.
\ea
The unprimed and primed sectors have their own horizontal (family) gauge groups
$SU(4)_H$ and $SU(4)_H^{\prime}$ respectively.

It would be amusing if the above group
comes from 
\ba
\label{tri}
&&SO(10)_{PS} \otimes SO(10) \otimes SO(10)^{\prime}
\rightarrow \nonumber \\
&&SU(4)_{PS} \otimes SO(10) \otimes SO(10)^{\prime}\,. 
\ea

The above presentation is in a nutshell the essence of the emergence of family
replication from spinors.
 
\item {\bf What makes the fourth neutrino much heavier than the other three?}

The Z-boson width constrains the fourth neutrino to be heavier than $M_Z/2$. It is natural then to ask
why this should be the case if a fourth generation exists. Although we do not have an answer
to that question, we will give a sketch of one scenario where one could
perhaps try as a first step toward finding that answer.

We will concentrate solely on the lepton sector in this section. The
purpose is to obtain the ratio of the Dirac masses $m_{D}^{(3)}/m_{D}^{(4)} \sim 10^{-6}$ 
for the active sector and $m_{D}^{(3),N}/m_{D}^{(4),N} \sim 10^{-4}$ for the
sterile sector according to the numerics discussed previously.
Let us introduce under $SU(4)_{PS} \otimes SU(4)_H \otimes SU(2)_L \otimes SU(2)_R
\otimes SU(4)_H^{\prime} \otimes SU(2)_L^{\prime} \otimes SU(2)_R^{\prime}$
\be
\label{horizontals}
\Phi_{HS} = (1,1,2,2,1,1,1)\,,
\ee
\be
\label{horizontala}
\Phi_{HA} = (1,15,2,2,1,1,1)\,.
\ee
The Dirac mass term similar to (\ref{yuk1}) can now be written, for the leptons, as
\be
\label{yukh}
{\cal L}_H= \bar{l}_L (g_S \Phi_{HS} + g_{S}^{A}\Phi_{HA})l^{M}_R + H.c. \,,
\ee
where $l_L$ and $l^{M}_R$ denotes a four-component (fundamental) representation of the
family $SU(4)_H$. (A term with exactly the same couplings is present for the
quarks.) With
\be
\label{vevphihs}
\langle \Phi_{HS} \rangle = v_S \,,
\ee
and
\be
\label{vevphiha}
\langle \Phi_{HA} \rangle = \frac{v_A}{2\sqrt{6}}
\left(\begin{array}{cccc}
1   &   &   &  \\
    & 1 &   &   \\
    &   &  1 &   \\
    &   &    & -3 
\end{array}\right) \,,
\ee
one readily obtain tree-level Dirac masses for the active neutrinos
\bes
\be
\label{3}
m_D = g_S v_S + g_{S}^{A}\frac{v_A}{2\sqrt{6}}\,,
\ee
\be
\label{4}
m_D^{(4)}= g_S v_S -3 g_{S}^{A}\frac{v_A}{2\sqrt{6}}\,,
\ee
\ees
for the common first three generations (\ref{3}) and the fourth
generation (\ref{4}) respectively. 

From (\ref{3}) and (\ref{4}), one first notices that $m_D \neq m_D^{(4)}$.
Because of this difference, one might ``fine-tune'' the VEVs so that
$g_{S}^{A}\frac{v_A}{2\sqrt{6}} \approx -g_S v_S$. If that can be done
then one could have a situation in which  $m_D \ll m_D^{(4)}$. A similar
consideration can be considered for the sterile neutrino sector. These
hints are under investigation.

It is beyond the scope of the paper to go deeper into this and related issues.
This will be presented elsewhere. To summarize, we have presented an argument outlining the
possibility of a heavy fourth neutrino, in addition to the three light ones. The main point
of the argument is simply the fact that there is {\em no} reason to expect additional
neutrinos to be as light as the SM neutrinos and that they violate the bound coming
from the Z width. 

\ei

\section{Conclusions}

A model with electroweak scale SM non-singlet right-handed neutrinos was presented
in \cite{hung1}. It contains a number of testable consequences such as lepton-number
violating processes at colliders through the direct production of the right-handed neutrinos
and their subsequent decays. In addition, there is a rich Higgs structure that can be
probed at colliders such as the upcoming LHC and the proposed ILC such as the existence of
doubly charged Higgs scalars contained in the model.

As shown in this paper, the attempt to unify quarks and leptons of the aforementioned model
in the manner of Pati-Salam fails unless one introduces new neutral fermions which are SM singlets,
the so-called sterile neutrinos. However, unlike generic models of sterile neutrinos where
they are usually thought of as SM-singlet right-handed particles, these new neutral fermions come
in {\em both} helicities:$N_L$ and $N_R$. This Pati-Salam extension basically ``completes'' the fermionic
assignments for the SM and mirror $SU(2)$ singlets: $(d_R,e_R)$, $(d^{M}_L, e^{M}_L)$,
$(u_R, N_R)$ and $(u^{M}_L,N_L)$. The gauge extension of the SM in this case is
the group $SU(4)_{PS} \otimes SU(2)_L \otimes
SU(2)_R \otimes SU(2)^{\prime}_L \otimes SU(2)^{\prime}_R$ with all the details given
in Section (\ref{putreview}). This group is reminiscent of the Petite Unification
model of \cite{put1} but differs from it in a major way, in terms of fermionic
assignments and patterns of symmetry breaking. The computation of $\sin^{2}\theta_{W}(M_Z)$
reveals the Pati-Salam scale to be of a typical GUT size. 
(In this sense, it has ``grown up''
and is no longer ``Petite''.) It is shown in Section (\ref{sterile}) that it is
reasonable to have keV sterile neutrinos which are only remotely constrained
by the active sector. These keV sterile neutrinos could constitute a part (or all?)
of warm dark matter and could be responsible for the so-called pulsar kicks.

The structure of the aforementioned gauge group and its fermionic
representations is very suggestive of the way spinors of some
orthogonal group decompose into spinors of its orthogonal subgroup.
For instance, a spinor of $SO(2m+4)$ decomposes into
$2^{m-1} \psi_{+} + 2^{m-1} \psi_{-}$ of $SO(4)$ or $2^{m-1}$ families
of $SO(4)$. With $SO(4) \approx SU(2) \otimes SU(2)$, it is argued in Section (\ref{spinor})
why the simplest, anomaly-free case where $m=3$ which corresponds to the group
$SO(10)$ and to {\em four} families is an appealing scenario. The group that
gives rise to this feature is argued to be $SU(4)_{PS} \otimes SO(10) \otimes SO(10)^{\prime}$
which breaks down to $SU(4)_{PS} \otimes SU(4)_H \otimes SU(2)_L \otimes SU(2)_R
\otimes SU(4)_H^{\prime} \otimes SU(2)_L^{\prime} \otimes SU(2)_R^{\prime}$ with
$SU(4)_H$ and $SU(4)_H^{\prime}$ being family gauge groups. It is argued
in Section (\ref{spinor}) how one might expect the fourth neutrino to be heavy.

\begin{acknowledgments}
I would like to thank Goran Senjanovic and Alexei Smirnov for
hospitality at the ICTP theory group  and for interesting
discussions while part of this work was carried out, and to Alex
Kusenko for useful communications. I would also
like to thank Alejandra Melfo and Borut Bajc for animated discussions
concerning the fourth family.
This work is supported
in parts by the US Department of Energy under grant No.
DE-A505-89ER40518.
\end{acknowledgments}

\end{document}